\documentclass[11pt]{article}
\usepackage{mymacros}
\usepackage{adjustbox}
\usepackage{tikz}
\usepackage{amsmath}
\usepackage{hyperref}
\usepackage{soul}
\usepackage{url}
\usetikzlibrary{arrows.meta, positioning, fit, backgrounds, calc, shapes.geometric, decorations.pathreplacing, calligraphy, decorations.markings}

\usetikzlibrary{shapes, arrows}

  {\begin{adjustbox}{minipage={#1}, margin=1cm 0.5cm}}%
  {\end{adjustbox}}

\title{Predictions with limited data: Bayesian (X)PINNs, entanglement surfaces and overconfidence}
\author[a]{Filip Landgren}
\author[b]{Marika Taylor}

\affiliation[a]{School of Mathematical Sciences and STAG Research Centre, University of Southampton, Highfield, Southampton SO17 1BJ, UK}
\affiliation[b]{College of Engineering and Physical Sciences
University of Birmingham, Birmingham, B15 2TT}
\emailAdd{f.landgrensoton.ac.uk, m.m.taylor@bham.ac.uk}
\abstract{
Solving differential equations from limited or noisy data remains a key challenge for physics-informed neural networks (PINNs), which are typically applied to already known and smooth solutions. 
In this work, we explore Bayesian PINNs and extended PINNs, (B-(X)PINNs), to solve non-linear second order differential equation typical for high energy theory, where data is only available from the boundary domain, to benchmark suitable approaches to PINNs in this category.
In particular, we consider an entangling surface; a differential equation typical in holography. We perform asymptotic analysis to generate analytical training data from the boundary domain. 
We also explore the meaning of overconfidence in models that are constrained by physical priors and argue that standard overconfidence metrics are not suitable to consider when dealing with B-PINNs. Overconfidence can be a natural feature and not a bug in systems with soft or hard constraints on the loss function; one have to look at when the overconfidence is an artifact of the model adhering to the physics constraints. To diagnose this effect, we introduce an information density quantity, and a local physics-constraint coupling (PCC) metric, to capture locally to what extent the enforced physics collapses the posterior distribution. We also consider these quantities for a Liouville-type equation and the Van der Pol equation to probe apparent overconfidence further. 
}

\begin{document}

\maketitle
\section{Introduction}
\begin{figure}[ht]
\centering
\begin{tikzpicture}[scale=0.9,
    roundnode/.style={circle, draw=black, very thick, minimum size=8mm},
    squarednode/.style={rectangle, draw=black, very thick, minimum size=6mm},
    every node/.style={align=center}
]

% Input Layer (light yellow fill)
\node[roundnode, fill=yellow!20] (I1) at (0,2) {};
\node[roundnode, fill=yellow!20] (I2) at (0,0) {};
\node[roundnode, fill=yellow!20] (I3) at (0,-2) {};
\node at (-0.8,2) {\small $x_1$};
\node at (-0.8,0) {\small $x_2$};
\node at (-0.8,-2) {\small $x_3$};

% Hidden Layer 1 (light blue fill)
\node[roundnode, fill=blue!20] (H11) at (3,3) {};
\node[roundnode, fill=blue!20] (H12) at (3,1) {};
\node[roundnode, fill=blue!20] (H13) at (3,-1) {};
\node[roundnode, fill=blue!20] (H14) at (3,-3) {};
\node at (3,3) {\small $h^{(1)}_1$};
\node at (3,1) {\small $h^{(1)}_2$};
\node at (3,-1) {\small $h^{(1)}_3$};
\node at (3,-3) {\small $h^{(1)}_4$};

% Hidden Layer 2 (light blue fill)
\node[roundnode, fill=blue!20] (H21) at (6,3) {};
\node[roundnode, fill=blue!20] (H22) at (6,1) {};
\node[roundnode, fill=blue!20] (H23) at (6,-1) {};
\node[roundnode, fill=blue!20] (H24) at (6,-3) {};
\node at (6,3) {\small $h^{(2)}_1$};
\node at (6,1) {\small $h^{(2)}_2$};
\node at (6,-1) {\small $h^{(2)}_3$};
\node at (6,-3) {\small $h^{(2)}_4$};

% Output Layer (light green fill)
\node[roundnode, fill=green!20] (O1) at (9,2) {};
\node[roundnode, fill=green!20] (O2) at (9,0) {};
\node[roundnode, fill=green!20] (O3) at (9,-2) {};
\node at (9.8,2) {\small $\hat{u}_1$};
\node at (9.8,0) {\small $\hat{u}_2$};
\node at (9.8,-2) {\small $\hat{u}_3$};

\foreach \i in {I1, I2, I3}
    \foreach \h in {H11, H12, H13, H14}
        \draw[->, thick] (\i) -- (\h);

\foreach \h in {H11, H12, H13, H14}
    \foreach \hh in {H21, H22, H23, H24}
        \draw[->, thick] (\h) -- (\hh);

\foreach \hh in {H21, H22, H23, H24}
    \foreach \o in {O1, O2, O3}
        \draw[->, thick] (\hh) -- (\o);

\draw[->, thick] (I1) -- (H11) node[midway, above, sloped] {\small $w^{(1)}_{11}$};
\draw[->, thick] (H11) -- (H21) node[midway, above, sloped] {\small $w^{(2)}_{11}$};
\draw[->, thick] (H21) -- (O1) node[midway, above, sloped] {\small $w^{(3)}_{11}$};

\node at (0, 3.5) {\small Input Layer};
\node at (3, 4) {\small Hidden Layer 1};
\node at (6, 4) {\small Hidden Layer 2};
\node at (9, 3.5) {\small Output Layer};

\end{tikzpicture}
\caption{Illustration of a two layer neural network where $x_i$ represent the input and $\hat{u}$ the predicted output.  $w^{(\ell)}_{ij}$ represents the weights connecting the neurons, $h^{(\ell)}_{i}$,  across layers.}
\label{fig:nn_architecture}
\end{figure}
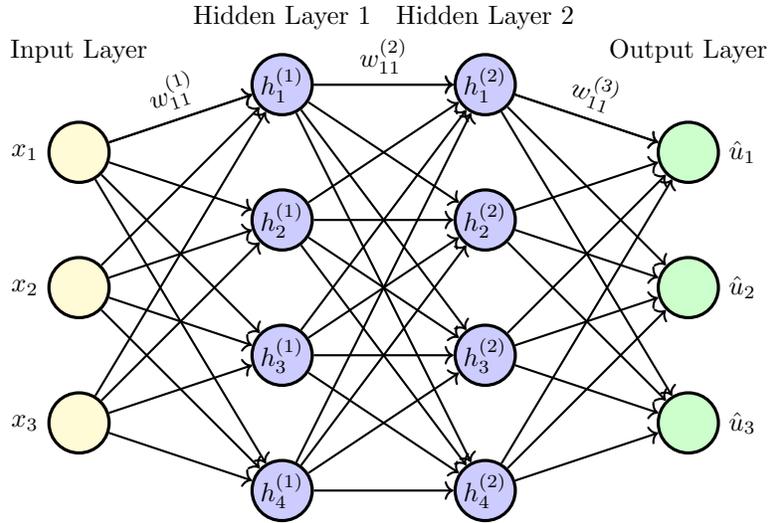

Machine learning and neural networks have been rapidly integrated into various domains in physics where data plays a crucial role \cite{rodrigues2023machinelearningphysicsshort}. Neural networks are promising for solving differential equations where traditional numerical methods fail, such as in the cases with high non-linearity. Their expressive power stems from their capacity to model non-linear relationships between inputs and outputs. Neural networks are purely data driven and learn from examples making connections with weights and biasses between modes to represent a function approximating the solution to the problem at hand, as illustrated in figure \ref{fig:nn_architecture}.  Physics-Informed Neural Networks (PINNs) \cite{RAISSI2019686}, introduces a "symbolic" element into the learning in terms of physical constraints in the loss function, typically in terms of penalizing deviations from boundary conditions and the residual. 

However, significant challenges remain when applying PINNs to problems where the solution is unknown or, for instance, where it is ill-behaved, non-unique or when training data is sparse \cite{zhang2024physicsinformedneuralnetworks}. In such cases, a naive PINN, without further guidance, may converge to an arbitrary solution branch.

Extended PINNs (XPINNs) augments ordinary PINNs by partitioning the domain into subdomains, each with its own separate network \cite{CiCP-28-2002}. This eases the learning in all sub-regions, and overall produces a better prediction, at the expense that the model is more prone to overfitting, due to potentially sparse data in the subdomains, and its inability to learn global features.  In \cite{Hu_2022}, the authors investigate how well XPINNs generalize, and use Barron space theory to find a trade-off condition when XPINNs generalize better than ordinary PINNs. XPINN's inability to learn global features is partly addressed by APINNs \cite{Hu_2023}, which allow flexible sharing of parameters between subnetworks, and by iPINNs \cite{dekhovich2023ipinns}, which learn incrementally by training each subnetwork sequentially, pruning over all previous subnetworks, and merging them into a single network.

Using XPINNs to solve ODEs and PDEs, with limited or noisy training data remains an active research area. In this work, we will focus on a complex ODE with two branches of solutions, with limited training data only near the domain boundaries. The data is multivalued, and
we will do a mild partitioning and let the model train on the two branches separately, but not divide the domains for each branch further. Since we are working with limited data, we will explore Bayesian physics-informed learning \cite{Yang_2021}, a B-XPINN, that uses stochastic learning and replaces the fixed weights in the network with (Gaussian) distributions. Through Bayesian inference, the model learns a posterior distribution over the network weights, which in turn induces a distribution over solutions. B-PINNs have proven particularly advantageous when working with limited and or noisy data \cite{Yang_2021, hou2024improvementbayesianpinntraining,Linka_2022}. Furthermore, the probabilistic treatment allows the model to quantify epistemic uncertainty from limited data, providing not just point estimates but also credible intervals for predictions. Such epistemic uncertainty estimates are crucial when working with sparse data, as they flag where the model is less certain and might benefit from additional data or refinement (see also e.g. \cite{ arbel2023primerbayesianneuralnetworks} for a review of Bayesian statistics in machine learning). 

We show that using domain decomposition and Bayesian inference, leads to more accurate and robust solutions compared to a standard PINN that lacks these features, when inferring the the solution from data only around the domain boundaries.

Central to this work is also the study of overconfidence and what it means for Bayesian physics-informed learning. Accurate measures for uncertainties were explored in \cite{kuleshov2018accurateuncertaintiesdeeplearning} and while the B-PINNs provides uncertainty estimates, interpreting and trusting these uncertainties requires care. An important question we investigate is how to ensure the model’s confidence is well-founded when it generalizes beyond the training region. In prediction tasks, a model is said to be overconfident if it estimates its uncertainty to be too low (or equivalently, is too certain in its predictions) in regions where it could actually be wrong. Overconfidence is a well-known issue in purely data-driven models, and often signals that the model is miscalibrated or overfit, failing to account for its lack of knowledge. In the context of physics-informed learning, however, the notion of overconfidence becomes more nuanced. A B-PINN heavily constrained by physical laws might appear overconfident even when it is correct, simply because the physical constraints eliminates degrees of freedom in the solution space. In other words, the model’s uncertainty can be very low not due to overconfidence in the usual sense, but because the physical prior confidently dictates the solution. It is thus crucial to distinguish between warranted confidence and misleading overconfidence in B-PINNs. In \cite{graf2022errorawareBPINNsimprovinguncertainty} it was recognized that conventional  B-PINNs merge measurement noise, parameter dispersion and equation error into a single posterior, masking the origin of the model’s certainty. They compensate by adding a pseudo-aleatoric variance term proportional to the PDE residual, which widens credible bands wherever the network violates the governing equation. Although this alleviates under-dispersion, it does not reveal why the model becomes confident, whether that confidence is earned from data or simply inherited from a physics prior. A parallel body of work has studied error propagation and coverage guarantees in PINNs \cite{shen2023picpropphysicsinformedconfidencepropagation,Bajaj_2023,daw2023mitigatingpropagationfailuresphysicsinformed,wu2025propinndemystifyingpropagationfailures}. These approaches tighten or calibrate prediction intervals, but they likewise leave unexplored the explicit contribution of the physics constraints to overconfidence.

Rather than treating all instances of high confidence as a flaw, regardless of origin, one should ask: when is the model’s confidence an artifact of limited data, and when is it a justified result of enforced physical laws? To diagnose overconfidence in physics-informed models, we introduce two metrics. The first is a gradient based information density measure (\ref{eq:sensitivity}), which assesses how much the observed data or physics constraints inform the posterior uncertainty of the model in different regions of the domain, by measuring sensitivity when varying the predicted output.

The second is a physics-constraint coupling (PCC) metric (\ref{eq:pcc gen}), which captures the degree to which the enforced physics constraints collapse the model’s posterior distribution. Moreover, the information density and local PCC evaluate how strongly the solution is determined by the physical prior relative to the data. A high local PCC can indicate that the physics conditions have tightly constrained the solution manifold, leaving little room for variation. By examining these metrics, we can better pinpoint regions where the model’s uncertainty is artificially low due to physics-driven constraints. A high confidence with low information density would raise a red flag, whereas regions with a high information density signals that the overconfidence is not necessarily bad and can even be expected.

The differential equation considered throughout this work is a non-linear second order ODE, corresponding to a non-trivial entangling surface on a negatively curved background. This is a typical differential equation in high energy theory, as thus serves as a good example to benchmark approaches to PINNs for these types of problems. The motivation also stems from the fact that the study of entangling surfaces and regions are typically restricted too smooth surfaces with low dimensionality \cite{Casini:2022rlv, Calabrese:2004eu}, and here we aim to make progress towards solving entangling surfaces with limited training data, that one can typically obtain with asymptotic analysis. 

An entangling surface is defined by the Euler–Lagrange equations one obtains when extremizing the area functional whose value computes the holographic entanglement entropy of a chosen boundary region. In static geometries, the Ryu-Takayanagi (RT) prescription \cite{Ryu:2006bv} picks out the co-dimension-2 minimal entangling surface. 
The extremality condition leads to a second-order, nonlinear PDE (or, under sufficient symmetry, an ODE) that admits closed-form solutions only in the simplest geometries, making these surfaces notoriously difficult to compute. 

Moreover, we will solve the annular entanglement surface considered in \cite{Landgren:2024ccz}, homologous to an annular entangling region in a three-dimensional negatively curved spacetime (AdS$_3$) residing on the boundary of AdS$_4$. 

Physics-informed learning has been widely utilized in engineering to address well-understood differential equations, such as those in fluid dynamics or heat transfer, where solutions are typically smooth and describe equilibrium or near-equilibrium states \cite{RAISSI2019686}. In contrast, high-energy physics problems, like the entangling surfaces explored in this work, generally involve non-smooth processes with complex behaviors, such as singularities and rapid gradient changes, common in quantum field theory and holography. The unpredictable nature of non-smooth or out-of-equilibrium high-energy physics pushes PINNs to their limits, requiring robust methods to ensure physically meaningful predictions; small parameter variations can lead to drastically different physical outcomes. The loss landscape of PINNs is in general not well understood \cite{rathore2024challengestrainingpinnsloss,Urb_n_2025}, which stems from the inherent difficulty of taking gradients of complicated differential equations; differential operators can even be ill-defined in certain domains. This complexity demands heightened caution when extending PINNs beyond the realm of well-behaved differential equations.

The remainder of this work is organized as follows: In section \ref{sec:pinn review}, we review physics-informed learning and in section \ref{sec: entangling surface}, we expand on entangling surfaces and present the ODE we are focusing on. In section \ref{sec:asymptotic analysis}, we generate the analytical training data from asymptotic analysis near the boundary. As a consistency check, we show that the divergent piece agrees with the covariant counterterm computed in \cite{Landgren:2024ccz}. In section \ref{sec:prepare model}, we prepare the model with numerical data and the boundary conditions. B-PINNs are reviewed in section \ref{sec:B-PINN}, where we also show the predicted solution of the entangling surface. In section \ref{sec:probing overconfidence} we diagnose overconfidence for the entangling surface and in section \ref{sec: further examples} we also consider the Liouville-type equation and the Van der Pol equation. Finally, we discuss our work in section \ref{sec:discussion}.

\subsection{Review of PINNs}\label{sec:pinn review}

Consider a network, $\N^{L+1}$, where $(L+1)$ is the number of layers, where the input layer is $\N^{0}(x)=x$. Each layer $\ell$ is represented by the weight matrix $W^{\ell} \in R^{M_{\ell-1}} \times R^{M_\ell}$ and the bias vector $\nu^{\ell}\in R^{M_\ell}$ where $M_{\ell}$ is the output size of $\N^\ell$. The output of each hidden layer is computed as (see, for instance, \cite{Hubeny:2012ry}):
\begin{equation}
\N^{\ell}(x)=\sigma \left( W^{\ell} \N^{\ell-1} (x)+\nu^\ell\right)
\end{equation}
where $\sigma$ is the activation function\footnote{popular choices include $\tanh(x), \text{ReLU}(x), \text{LeakyReLU}(x)$.}. The outputs in the final layer $L$ is given by
\begin{equation}
    \N^{L}(x)=\hat{u}_\theta(x)=W^{L} \N^{L-1}(x)+\nu^L=(\N^L \circ \N^{L-1} \cdots \N^{0})(x)
\end{equation}
where the last line is the sequence of non-linear functions and $\circ$ is the function composition and $\theta=\{W^{\ell}, \nu^\ell \}_{\ell=1, L}$ is the learning parameter, representing the weights or parameters of the model.

PINNs \cite{RAISSI2019686} enhance neural network training by incorporating underlying physical constraints directly into the loss function.
\begin{equation}
    L=w_i L_i
\end{equation}
where $L_i$ is any (normalized) physical constraint or information we have about the solution, and $w_i$ the corresponding weight. Consider, for instance, a general PDE of the form $\mathcal{N}[u(x)]=f(x)$ where $\mathcal{N}$ is some differential operator, with the boundary condition $\mathcal{B}[u(x)]=b(x)$ and the initial condition $u(x)=c(x)$. Let the predicted network output be denoted as $\hat{u}(\theta, x)$, then the total loss function takes the form 
\begin{equation}\label{eq: tot loss}
    L=L_\mathcal{N} + L_{u_0} + L_b
\end{equation}
where
\begin{align}
    \mathcal{L}_{\mathcal{N}}&=\frac{1}{N_\mathcal{N}} \sum_{i=1}^{N_{\mathcal{N}}}||\mathcal{N}[\hat{u}(\theta, x_i)]-f(x_i)||^2\\
    \mathcal{L}_{u_0}&=\frac{1}{N_{u_0}} \sum_{j=1} || \hat{u}(\theta, x_j)-c(x_j)||^2\\
    \mathcal{L}_b&=\frac{1}{N_b}\sum^{N_b}_{k=1}||\mathcal{B}[\hat{u}(\theta, x_k)]-b(x_k)||^2.
\end{align}
Here, $N_{\mathcal{N}}$ represents the number of points used to fit the predictions of the neural network to the observed data. $N_{u_0}$ and $N_b$ represent the number of collocation points where the initial and boundary conditions are enforced, respectively. We may add more constraints other than initial conditions, boundary conditions and the residual, such as enforcing the solution or gradient values at more points, monotonicity conditions or any other insights from the the solution.

The neural network is now physics-informed through effective regularisation in the sense that deviations from the initial conditions, boundary conditions as well as the residual of the physical system, are penalized during learning as we minimize the loss function with respect to the learning parameter $\theta$. 
$\theta$ will not appear explicitly in the neural network as it is implicitly represented by the weights. As the network updates the model parameters to minimize the loss function during training, the weights are computed recursively:
\begin{equation}
    \theta^{j+1}=\theta^j-l_r\nabla_\theta L(\theta^j)
\end{equation}
where $L$ is the $j$-th iteration that we call an epoch and $l_r$ is the learning rate.  At its core, PINNs computes gradients with the chain rule. Although the idea of PINNs have been around since the 80s, they have only been practical since the development of libraries such as PyTorch and TensorFlow, making automatic differentiation to compute $\nabla_\theta$ more tractable. 
In this work we use PyTorch, due to its versatile nature, combining ease of use with powerful modules.

Ordinary PINNs or "vanilla PINNs" have been useful for solving a host of  differential equations ranging from Helmholtz equations to Laplace equations \cite{baty2023solvingstiffordinarydifferential, baty2024handsonintroductionphysicsinformedneural, LEE1990110, MEADE199419,Yentis1996VLSIIO} (see also \cite{cuomo2022scientific} for a review). While PINNs are cutting edge methods of obtaining a solution to a differential equation, their naive application is sensitive to numerical instabilities. In particular, cases with high-frequency behaviors, casps, sudden steep changes in the gradients, or multi-valued data, for instance, can quickly cause their performance to deteriorate; training is hindered by the complexity and non-convexity of the loss function. In this work we investigate the optimal approach to PINNs for typical holography equations.

\begin{figure}[ht]
\centering
\begin{tikzpicture}[scale=0.9, every node/.style={transform shape},
    neuron/.style={circle, draw, minimum size=1cm, inner sep=0pt},
    layer/.style={rectangle, draw, rounded corners, minimum height=4cm, minimum width=1.5cm, fill=none},
    arrow/.style={-Stealth, thick},
    label/.style={font=\footnotesize}
]

\node[neuron, fill=yellow!20] (x) at (0, 1) {$x$};
\node[neuron, fill=yellow!20] (t) at (0, -1) {$t$};
\node[label, above=0.5cm] at (x.north) {Inputs};

\node[layer] (h1) at (3, 0) {};
\node[label, above=0.5cm] at (h1.north) {Hidden Layer 1};
\foreach \i in {-1.5, -0.5, 0.5, 1.5} {
    \node[neuron, fill=blue!20] at (3, \i) {};
}

\node[layer] (h2) at (6, 0) {};
\node[label, above=0.5cm] at (h2.north) {Hidden Layer 2};
\foreach \i in {-1.5, -0.5, 0.5, 1.5} {
    \node[neuron, fill=blue!20] at (6, \i) {};
}

\node[neuron, fill=green!20] (u) at (9, 0) {$\hat{u}(x,t)$};
\node[label, above=0.5cm] at (u.north) {Output};

\node[rectangle, draw, rounded corners, minimum height=2cm, minimum width=3cm, fill=blue!10] (loss) at (12, -2) {Physics Loss: \ $\nabla_\theta L \to 0$};
\node[label, above=0.5cm] at (loss.north) {PDE Constraint};

\draw[arrow] (x) -- (h1);
\draw[arrow] (t) -- (h1);
\foreach \i in {-1.5, -0.5, 0.5, 1.5} {
    \draw[arrow] (h1) -- (6, \i);
}
\draw[arrow] (h2) -- (u);
\draw[arrow] (u) -- (loss);

\draw[arrow, dashed] (loss.west) to[out=180, in=-90] (9, -3) to[out=90, in=-90] (h2.south);

\end{tikzpicture}
\caption{Schematic sketch of a PINN architecture illustrating that the connections are made such that the loss function, with any underlying PDE constraints, is minimized. }
\label{fig:pinn_colored}
\end{figure}
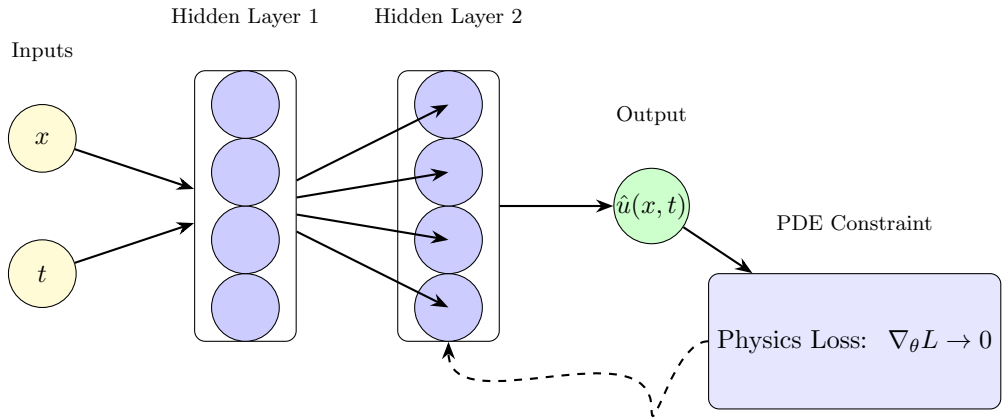

\section{Entangling surfaces}\label{sec: entangling surface}
As a illustrative example of a typical ODE that arises in high-energy theory and holography, we consider an entangling surface. Entangling surfaces generally satisfy highly non-linear PDEs, but in the symmetric setup we consider the problem reduces to an ODE, which remains notoriously difficult to solve. Consequently, most studies restrict attention to cases with simple, smooth symmetries \cite{Landgren:2024ccz}.

In the Anti–de Sitter / conformal field theory correspondence (AdS/CFT) \cite{Maldacena:1997re}, an entangling surface is the co-dimension-2 hypersurface in the bulk that extremizes an area functional and is homologous to a given boundary region \cite{Taylor:2016aoi} of the entangling region in question. In static spacetimes \cite{Hubeny:2007xt} this extremal surface is known as the Ryu–Takayanagi (RT) surface \cite{Ryu:2006bv}.

The holographic entanglement entropy of a static entangling region $A$ of a CFT$_d$, with an asymptotically AdS$_{d+1}$ dual, is given in terms of the area of the the ($d-1$)-dimensional RT surface, $\gamma$, with $\partial \gamma = \partial A$, as:
\begin{equation}
S_{\text{vN}}=\frac{\mathcal{A}_A[\gamma_\epsilon]}{4G_N}
\end{equation}
where $\mathcal{A}_A[\gamma_\epsilon]$ is the area of the regularized co-dimension two hypersurface $\gamma_\epsilon$ and  $G_N$ is the ($d+1$)-dimensional Newton's constant. 

More generally, the study of (dynamical) surfaces governed by an area functional has applications throughout physics, such as in fluid phases, small deformations of elastic membranes at the mesoscopic scale, cosmic strings \cite{carter1997branedynamicstreatmentcosmic}, the coupling between quantum field theories and defects \cite{Johnson:2000ch, Jensen:2015swa,McAvity:1993ue}, and D-brane dynamics \cite{Bachas:1995kx}, just to name a few. 
In this work, we will focus on entangling surfaces situated on a hypersurface of constant time, although we in principle could consider time dependence by evaluating the Hubeny-Rangamani-Takayanagi (HRT) surfaces \cite{Hubeny:2007xt}, the covariant counterpart to RT surfaces. 

For an entangling surface $A$ in asymptotically AdS$_4$ spacetimes, the  entanglement entropy can be written as \cite{Fonda:2015nma}
\begin{equation}\label{eq: entropy with F term}
    \A[\gamma_\epsilon]=c_{-1}\frac{\mathcal{L}}{\epsilon}+c_0+\ldots
\end{equation}
where $c_{-1}$ and $c_0$ are dimensional constants and $\mathcal{L}$ is the length of the boundary and requires complete knowledge of the entangling surface. In condensed matter theory, where $\epsilon$ is a lattice cutoff, $c_{-1}$ might be physical, whereas in QFT $\epsilon$ just serves as a regulator. In the limit $\epsilon \to 0$, $c_0$ is the first non-trivial term depending on the entire entangling surface allowing the IR geometry to be probed for a sufficiently large entangling region \cite{Fonda:2015nma}.
For finite entangling regions, the analytical expressions of the $c_0$ term is known only in symmetrical cases like that of a disk \cite{Taylor:2016aoi} or an annulus, on flat backgrounds (see, for instance, \cite{Drukker_2006}).
In the limit $\epsilon \to 0$, the shape dependence of higher order terms in the entanglement entropy has been widely studied (see e.g. \cite{Fonda:2015nma, Solodukhin:2008dh,Klebanov:2012yf, Hubeny:2012ry, Astaneh:2014uba, Fonda:2014cca}).  Even obtaining numerical solution often poses a great challenge. 
In \cite{Fonda:2015nma}, a closed-form expression for $c_0$ was obtained for a finite entangling region, in an asymptotically AdS$_4$ bulk spacetimes whose boundary is a three-dimensional Minkowski spacetime boundary, using the Willmore energy formula \cite{willmore1965note}  for the minimal surface. 
The authors of \cite{Fonda:2015nma} used the Surface Evolver program\footnote{ The Surface Evolver program was built to generally understand energy-minimizing surfaces, and was first applied in the context of holography and entropy in \cite{DeNobili:2015dla} to better understand the shape dependence of holographic mutual information.}\cite{brakke1992surface}, to numerically compute the entangling surface to cross-check their results.

In the study of entanglement entropy from the QFT side, Monte Carlo simulations, machine learning, and deep learning techniques have primarily been applied to condensed matter lattice systems (see, for example, \cite{Bulgarelli:2024yrz, Humeniuk_2012, Luitz_2014, PhysRevX.7.021021, Inglis_2013}).
In holographic setups, particularly within the context of AdS/CFT, machine learning has largely been utilized for reconstructing isotropic bulk spacetimes given a dual quantum field theory and corresponding entanglement entropy data \cite{Park_2022, Ahn:2024lkh, Song:2020agw}. However, the application of machine learning to directly solve for entangling surfaces or holographic entanglement entropy remains largely unexplored.

In this work, we will use Bayesian physics-informed learning towards solving the extrimization problem, in terms of minimizing the area functional representing the entanglement entropy on non-trivial curved backgrounds. Furthermore, we will consider the entangling surface of an annular entangling region in AdS$_3$, residing on the non-compact boundary of AdS$_4$\footnote{Since only one scale in the problem, the width of the annulus, with the rest of the dimensions being isometric circular directions, the study can straight forwardly be generalized to arbitrarily dimensions. For more details on this see \cite{Landgren:2024ccz}.}, studied in \cite{Landgren:2024ccz} which we summarize below. 

This annular setup provides a nontrivial benchmark for our Bayesian physics-informed learning approach: the minimal surface equation admits no known closed-form solution because the curved background, which brakes  translational invariance and symmetry about the inflection point. In \cite{Landgren:2024ccz}, the entanglement entropy was obtained indirectly, via a flat-space limit of the holographic construction, circumventing a direct solution of the governing ODE. Despite this complexity, the resulting entangling surface is expected to be smooth, without cusps or singularities.  Since only one physical scale appears, the annulus width, with all other directions being isometric, the analysis generalizes straightforwardly to higher dimensions, and the governing PDE simplifies to an ODE.

We will construct our model to function with limited minimal training data, namely analytical data from asymptotic analysis near the conformal boundary, supplemented with a small sample of numerical data around the inflection point, and infer the solution in the intermediate data-absent regions. Challenging features of our solutions are multi-valued data, large gradient values, and a tightly confined domain and range. 
We now proceed to the setup of the differential equation to be analyzed. The AdS$_4$ geometry can be described in terms of the C-metric.
The AdS$_4$ C‐metric describes two black holes accelerating in opposite directions under the tension of a cosmic string that threads the wormhole between them.  This string introduces conical singularities into the global geometry, so any RT surface must avoid plunging too deeply into the bulk to remain causally disconnected from those singularities.  By choosing a sufficiently small boundary region, one ensures the corresponding extremal surface stays close to the AdS boundary.  
In entanglement island constructions \cite{Almheiri:2019yqk,Penington:2019kki}, one endpoint of the RT surface is anchored to the boundary while the other is fixed by the island rule, which might in principle pull the surface deeper into the bulk.  However, as suggested in \cite{Landgren:2024ccz}, even in that setup the extremal surface does not venture far enough to encounter the conical singularities. Although the precise effects of causal contact with the singularities remain unclear, any resulting discrepancies should become apparent in the calculation.

In global coordinates the C-metric can be expressed as: 
\begin{equation}\label{eq: ads4}
    ds_4^2 = \ell_4^2 d\sigma^2 + \frac{\ell_4^2}{\ell_3^2}\cosh^2\sigma \left(  \frac{dr^2}{\frac{r^2}{\ell_3^2} + \kappa } - \left(\frac{r^2}{\ell_3^2} + \kappa \right)dt^2  + \phi_c^2d\tilde{y}^2\right).
\end{equation}
In these coordinates, the conformal boundary is located at $\sigma \to \infty$. 
On transforming the conformal AdS$_3$ boundary from global to Poincaré coordinates we have
\begin{equation}\label{Poincarémet}
    ds_4^2 = d\sigma^2 \ell_4^2 + \ell_4^2 \cosh ^2\sigma  \left( \frac{dx^2-dt^2}{x^2} + \frac{\phi_c^2 dy^2 }{x^2}\right).
\end{equation}
The boundary metric (at $\sigma \to \infty$) is the uplifted AdS$_2$ metric \cite{Landgren:2024ccz}:
\begin{equation}\label{eq:Poincaré boundary}
     ds_3^2 = \ell_4^2 \left( \frac{dx^2-dt^2}{x^2} + \frac{\phi_c^2 dy^2 }{x^2} \right).
\end{equation}

By parameterizing the RT surface with worldvolume coordinates $x^\alpha =\{\sigma, y\}$, with the embedding coordinates  $x^m=\{t, \sigma, x(\sigma), y\}$, the area functional for the regulated entropy becomes 
\begin{equation}\label{Sxsigma}
    S_{\textnormal{reg}}= \frac{1}{4 G_4}\int_0^{2 \pi} dy \left( \int_{\frac{1}{\epsilon}}^{\sigma_0}  d \sigma   \mathcal{L}\left((x_{b}(\sigma), x_{b}'(\sigma), \sigma \right) +   \int_{\sigma_0}^{\frac{1}{\epsilon}} d \sigma    \mathcal{L}\left((x_{a}(\sigma), x_{a}'(\sigma), \sigma \right) \right)
\end{equation}
where 
\begin{equation}
   \mathcal{L}\left(x(\sigma), x'(\sigma), \sigma \right)= \frac{\ell_4^2 \phi_c \cosh \sigma}{x(\sigma )}\sqrt{\frac{\cosh^2 \sigma x'(\sigma
   )^2}{x(\sigma )^2}+1}.
\end{equation}
As noted above, the RT surface lacks reflection symmetry about its inflection point, so the equations of motion yield two distinct solution branches, $x_a(\sigma)$, $x_b(\sigma)$.

The area functional (\ref{Sxsigma}) is extrimized by solving the differential equation
\begin{multline}\label{eq: diff sigma}
   \cosh (\sigma ) x(\sigma )^2
   \left(\cosh (\sigma ) x''(\sigma )+3 \sinh (\sigma ) x'(\sigma )\right) \\+ 2 \sinh (\sigma ) \cosh ^3(\sigma ) x'(\sigma )^3 +x(\sigma)^3=0.
\end{multline}
The RT surface is the solution $x(\sigma)$ that has a turning point at $(x_0,\sigma_0)$ in the bulk and intersects the boundary at $(\sigma \to \infty,x_1)$ and $(\sigma \to \infty,x_2=x_1+L)$. We expect two branches of solution corresponding to whether the solution intersects the boundary at $x_1$ or $x_2$: $x_a(\sigma)$ and $x_b(\sigma)$. Hence, we have the boundary conditions
\begin{align}
    x_a(\infty)&=x_1, \quad x_b(\infty)=x_2\\
    x_a(\sigma_0)&=x_b(\sigma_0)=x_0\\
    x_a'(\sigma_0)&=x_b'(\sigma_0)=\infty  .
\end{align}
Carrying out a change of coordinates $\xi=e^{-2\sigma}$, we can write (\ref{eq: diff sigma}) as
\begin{multline}\label{xdeqn}
    (\xi-1) u (\xi+1)^3 x'(\xi)^3+\frac{1}{2} (\xi+1) x(\xi)^2 \big(2 \xi (\xi+1) x''(\xi)\\+(5 \xi-1)
   x'(\xi)\big)+x(\xi)^3=0
\end{multline}
where the conformal boundary is now at $\xi=0$. Further changing coordinates to $x(\xi)=e^{f(\xi)}$ we get,
\begin{multline}
    \xi (\xi+1)^2 f''(\xi)+\frac{1}{2} \left(5 \xi^2+4 \xi-1\right) f'(\xi)+(\xi-1) \xi (\xi+1)^3
   f'(\xi)^3 \\+\xi (\xi+1)^2 f'(\xi)^2+1=0.
\end{multline}
We can immediately notice that the resulting differential equation depends only on $f''(\xi)$ and $f'(\xi)$. Hence we can now split the second-order ODE into two first-order ODEs:
\begin{align} 
    &f'(\xi)=g(\xi)\label{fdeqn}\\
   \begin{split} &\xi (\xi+1)^2 g'(\xi)+\frac{1}{2} \left(5 \xi^2+4 \xi-1\right) g(\xi)+(\xi-1) \xi (\xi+1)^3 g(\xi)^3\\
   &+\xi(\xi+1)^2 g(\xi)^2+1=0. 
   \end{split} \label{eq: gdeqn}
\end{align}
Equivalently, (\ref{eq: diff sigma}) can be written as
\begin{align}\label{eq:diff u(x)}
   \begin{split} &4\xi(x)^4 -2\xi(x)^5+x^2 \xi'(x)^2(1-2x\xi'(x))+2x^2\xi(x)^3 \xi''(x)\\
   &+\xi(x)\left(2-4x^2 \xi'(x)^2 +2x^2 \xi''(x) \right) + \xi(x)^2 \left(4-5x^2 \xi'(x)^2 +4x^2 \xi''(x) \right)=0 
   \end{split}
\end{align}
using
\begin{equation}
    \sigma'(x)=-\frac{\xi'(x)}{2\xi(x)}, \quad \sigma''(x)=\frac{1}{2}\left( \frac{\xi'(x)^2}{\xi(x)^2}-\frac{\xi''(x)}{u(x)} \right).
\end{equation}

At the point $\xi=0$, we have from (\ref{eq: gdeqn}) that  
\begin{align}
    &g(0)=2\\
    & x'(0)=2x(\xi=0)=2x_{1,2}
\end{align}
where $x_{1,2}$ are the endpoints at the conformal boundary where the RT surface is homologous to the entangling region. The function $g(\xi)$ determines $x(\xi)$ up to some overall scaling i.e., $\, x(\xi, \xi_0, x_0)=\lambda x(\xi, \xi_0, \lambda x_0)$. Furthermore, at the inflection point, we observe from (\ref{eq: gdeqn}) evaluated at the inflection point $\xi_0$ that the range of the surface is bounded by $0<\xi_0<1$ from the fact that $g'(\xi_0)\to \infty$ if $g(\xi_0) \to \infty$.
We will use asymptotic analysis around the boundary $\sigma \to \infty$ to generate training data near the conformal boundary, to feed the deep networks.

\subsection{Asymptotic analysis}\label{sec:asymptotic analysis}

Solving (\ref{eq: gdeqn}) we get the implicit relation for $g(\xi)$
\begin{equation}\label{gimplicit}
    \frac{\sqrt{\frac{\xi-1}{\xi}} \left(\frac{(2 (\xi+1) \xi g(\xi)-\xi+1) \, _2F_1\left(\frac{1}{4},1;\frac{3}{2};-\frac{(-2 (\xi+1) g(\xi) \xi+\xi-1)^2}{\xi \left(\left(\xi^2-1\right)
   g(\xi)+2\right)^2}\right)}{\left(\xi^2-1\right) g(\xi)+2}+\xi-1\right)}{2 \sqrt{1-\xi} \sqrt[4]{-\frac{(-2 (\xi+1) \xi g(\xi)+\xi-1)^2}{\xi \left(\left(\xi^2-1\right) g(\xi)+2\right)^2}-1}}=C_1
\end{equation}
where $C_1$ is the integration constant. Reinstating the coordinates $x(\xi)$ we have 
\begin{equation}\label{gxrln}
    g(\xi)=f'(\xi)=\frac{\partial(\log[x(\xi)])}{\partial u}=\frac{x'(\xi)}{x(\xi)}
\end{equation}
Substituting this back in (\ref{gimplicit}) and imposing the boundary condition at the turning point $x'(\xi_0=e^{-2\sigma_0})=\infty$ we fix $C_1$ in terms of $\xi_0=e^{-2\sigma_0}$    :  
\begin{equation}\label{C1}
    C_1(\xi_0)=-\frac{\sqrt{\frac{\xi_0-1}{\xi_0}} \left(2 \xi_0 \, _2F_1\left(\frac{1}{4},1;\frac{3}{2};-\frac{4 \xi_0}{(\xi_0-1)^2}\right)+(\xi_0-1)^2\right)}{2
   (1-\xi_0)^{3/2} \sqrt[4]{-\frac{(\xi_0+1)^2}{(\xi_0-1)^2}}}
\end{equation}
encoding information about the turning point.
Now considering (\ref{gimplicit}) and (\ref{gxrln}), we have the general relation
\begin{equation}
    g(\xi)=\frac{x'(\xi)}{x(\xi)}= P(\xi,C_1(\xi_0)) 
\end{equation}
for a general function $ P(\xi,C_1(\xi_0))$. Solving for $x(\xi)$ gives us
\begin{equation}\label{xgrln}
    x(\xi)=C_2e^{\int du P(\xi,C_1)}
\end{equation}
where $C_2$ is the second integration constant that acts as an overall scaling. This can also be observed from the differential equation for $x(\xi)$ (\ref{xdeqn}) where we see that $C_2 x(\xi)$ is a solution if $x(\xi)$ is a solution. We see that the asymptotic analysis of $\xi\to0$  shows that $e^{\int du P(\xi,C_1)} \to 1$ as $\xi\to 0$.

Now, imposing the boundary condition $x(0)=x_1, x_2$ along with $x'(\xi_0)=\infty$, we get two branches of solutions, one with $C_2=x_1$ and the other with $C_2=x_2$.  $C_2$ is independent of the choice of $C_1$. In other words, $C_2$ only captures where the curve intersects the boundary and is independent of $C_1$ which only captures information about the turning point $\xi_0$. 

Close to the boundary, we can write down the following ansatz for a particular $g(\xi)$:
   \begin{equation}
    g(\xi) = \sum_{n=0}^{\infty}a_n \xi^n.
\end{equation}
Using this ansatz and solving perturbatively order by order for $a_n$ we get
\begin{equation}\label{gpart}
    g(\xi) = \sum_{n=0}^{\infty}2 \xi^{2n}=\frac{2}{1-\xi^2}.
\end{equation}
This is a particular solution for $g(\xi)$. Reinstating the coordinates $x(\xi)=e^{\int d\xi g(\xi)}$ we get a one-parameter family of solutions for $x(\xi)$
\begin{equation}\label{opx}
    x(\xi) = C_3 \left(\frac{1+\xi}{1-\xi}\right).
\end{equation}
From our previous analysis of the full solution for $x(\xi)$ we see that this particular solution corresponds to a choice of the integration constant $C_1(\xi_0)$. $C_3$ in this particular solution is the scaling constant.  Since $C_3$ is independent of $C_1$, we could plug in the derivative of the particular solution for $x(\xi)$ (\ref{opx}) into (\ref{gxrln}) and (\ref{gimplicit}), to get an implicit full solution for $x(\xi)$. 
Combining this with the results we got for $C_1(\xi_0)$ we get,
\begin{multline}
   \frac{\sqrt{\frac{\xi-1}{\xi}} \left(\frac{\left(\frac{4 C_3 (\xi+1) \xi}{(\xi-1)^2
   x(\xi)}-\xi+1\right) \, _2F_1\left(\frac{1}{4},1;\frac{3}{2};-\frac{\left(-\frac{4 (\xi+1) C_3 \xi}{(\xi-1)^2 x(\xi)}+\xi-1\right){}^2}{\xi \left(\frac{2 \left(\xi^2-1\right) C_3}{(\xi-1)^2
   x(\xi)}+2\right){}^2}\right)}{\frac{2 C_3 \left(\xi^2-1\right)}{(\xi-1)^2 x(\xi)}+2}+\xi-1\right)}{2 \sqrt{1-\xi} \sqrt[4]{-\frac{\left(-\frac{4 C_3 (\xi+1) \xi}{(\xi-1)^2 x(\xi)}+\xi-1\right){}^2}{\xi
   \left(\frac{2 C_3 \left(\xi^2-1\right)}{(\xi-1)^2 x(\xi)}+2\right){}^2}-1}}\\=C_1(\xi_0)= -\frac{\sqrt{\frac{\xi_0-1}{\xi_0}} \left(2 \xi_0 \, _2F_1\left(\frac{1}{4},1;\frac{3}{2};-\frac{4 \xi_0}{\left(\xi_0-1\right){}^2}\right)+\left(\xi_0-1\right){}^2\right)}{2
   \left(1-\xi_0\right){}^{3/2} \sqrt[4]{-\frac{\left(\xi_0+1\right){}^2}{\left(\xi_0-1\right){}^2}}}.
\end{multline}
This implicit solution for $x(\xi)$ is still difficult to unpack and we will instead analyze the behavior close to the boundary. 

Consider expanding the particular solution (\ref{gpart}) near the boundary. 
\newline Since $0<\xi\leq \xi_0 <1$ a natural expansion parameter for a perturbative series is any function $f(\xi)$ such that $0<f(\xi)<1$. We choose the expansion parameter $f(\xi)=q=\sqrt{\xi}$ and consider an ansatz for $g(\xi)$ of the form
\begin{equation}
    g(\xi)= \frac{2}{1-\xi^2}+q \sum_{n=0}^{\text{order}}h_n q^n.
\end{equation}
We can plug this ansatz into the differential equation for $g(\xi)$ and solve for $h_n$ order by order perturbatively. We have listed $h_n$ up to $h_6$ below: 

\begin{align}
    h_0 &= k,\\
    h_1 &= 0,\\
    h_2 &= 5k,\\
    h_3 &= (10 k^2)/3,\\
    h_4 &= k (28 + k^2)/2,\\
    h_5 &= (80 k^2)/3,\\
    h_6 &= 30 k + (305 k^3)/18
\end{align}
where $k$ is the integration constant. Reinstating the coordinates $x(\xi)=e^{\int du g(\xi)}$ we get,
\begin{equation}\label{asymx}
    x(\xi; k, C_2) = C_2 \frac{1+\xi}{1-\xi} e^{\frac{2}{3} k \xi^{3/2}\left(1+ 3 \sum \limits_{n=2}^{\text{order}} \frac{h_n}{k} \xi^{n/2} \right)}
\end{equation}

\begin{figure}[ht]
    \centering
    \includegraphics[width=0.7\linewidth]{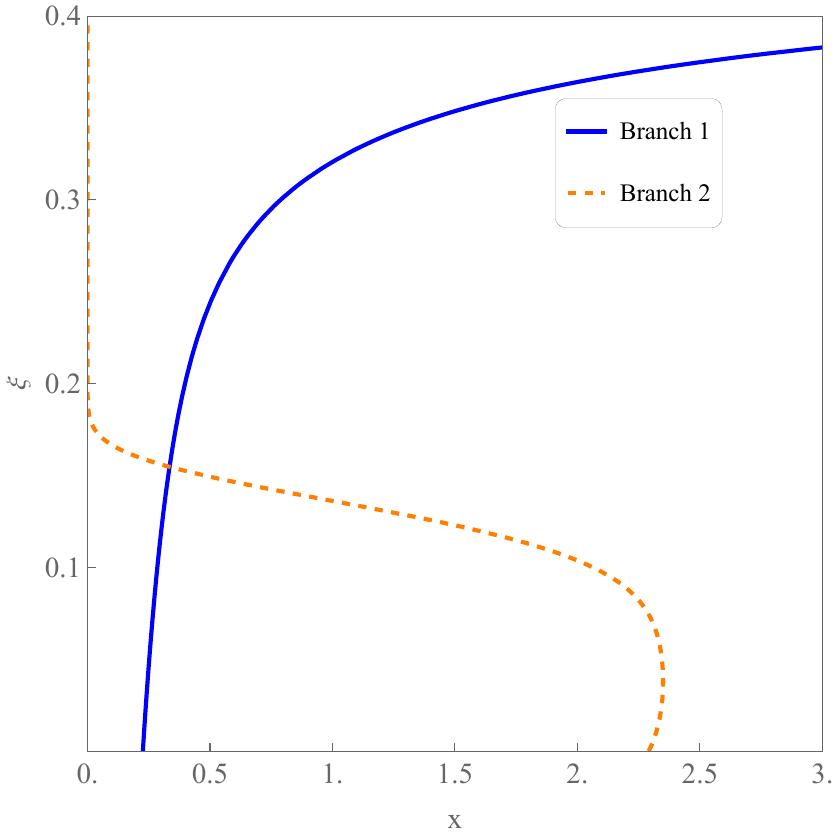}
    \caption{Plot of analytical data for two values for $C_2$ and $k$ for the two branches of solutions. As expected this data is only accurate for small $\xi$ near the conformal boundary $\xi=0$.}
    \label{fig:analytical data}
\end{figure}
In Figure \ref{fig:analytical data}, we plot the analytical asymptotic solution, which is reliable only close to the boundary at $\xi=0$. As one moves away from the boundary, the curve rapidly departs from the true behavior, signaling the breakdown of the asymptotic approximation.We see that there is a turning point for the yellow curve where the derivative switches from negative to positive rendering $\xi(x)$ multi-valued.
Since $C_2$ is just the scaling constant and $x(\xi) \to C_2$ as $\xi\to0$, therefore $C_2 = x_1, x_2$. These two choices along with corresponding choices for the constant $k=k_1,k_2$ gives two branches of solutions, $x_a(\xi;x_1, k_1)$ and $x_b(\xi;x_2, k_2)$, on which the matching boundary conditions at the turning point have to be imposed to fix $ k_1(x_1,x_2)$ and $k_2(x_1,x_2)$.

The divergent contributions to the area functional (\ref{Sxsigma}) originate near the boundary. To isolate and extract these divergences, we consider the asymptotic expansion of 
$x(\xi)$ around the boundary, retaining terms up to the order necessary to capture the complete divergent structure. In $x(\xi)$ coordinates the area functional (\ref{Sxsigma}) takes the form
\begin{equation}\label{sxu}
    S_{\textnormal{reg}} = \frac{1}{4G_4}\int _0^{2\pi} dy \left(\int_{\epsilon}^{\xi_0} d u \mathcal{L}(\xi, x_a(\xi;k_1,x_1)) + \int_{\xi_0}^{\epsilon} d\xi \mathcal{L}(\xi; x_b(u;k_2,x_2)) \right)
\end{equation}
with 
\begin{equation}
\mathcal{L}(\xi) = \frac{-1}{4 \xi} \sqrt{\frac{\ell_4^4 (\xi+1)^2 \phi_c ^2 \left(\xi (\xi+1)^2 x'(\xi)^2+x(\xi)^2\right)}{\xi x(\xi)^4}}
\end{equation}
where $x_a(\xi;k_1,x_1)$ and $x_b(\xi;k_2,x_2')$ are the two branches intersecting the boundary at $x_1, x_2$ respectively.

Substituting the asymptotic series solution of $x(u; k,C_2)$ around the boundary (\ref{asymx}) into $\mathcal{L}(\xi)$, and expanding around $\xi=0$ gives
\begin{equation}
    \mathcal{L}(\xi;k, C_2) = \frac{\phi_c \ell_4^2}{C_2} \left( \frac{-1}{4 \xi^{3/2}}-\frac{1}{4 \xi^{1/2}}-\frac{k}{3}-\frac{k^2}{8} \xi^{1/2}-\frac{4y}{3} \xi-\frac{125 k^2}{72} \xi^{3/2} \right) + \mathcal{O} (\xi^2). 
\end{equation}
Only the first term $\frac{\phi \ell_4^2}{C_2} \left( \frac{-1}{4 \xi^{3/2}}\right)$ in $\mathcal{L}(\xi;k, C_2)$ contributes to the divergence in the entanglement entropy. Since we are considering the series solution of $x(\xi)$ around the boundary from where the divergent contributions reside, more terms in the asymptotic series for $x(\xi)$ will not give additional contributions to the divergence. 

Plugging $\mathcal{L}(\xi)$ back into the entropy functional (\ref{sxu}), we get the divergent contribution to the entanglement entropy in full generality given by
\begin{equation}\label{Sdivergent}
    S_{\text{div}} =  \frac{\pi\phi_c \ell_4^2}{4G_4 \sqrt{\epsilon}}\left(\frac{1}{x_2} - \frac{1}{x_1}\right)
\end{equation}
which completely agrees with covariant counterterm computed in \cite{Landgren:2024ccz} derived with the formula
\begin{equation}
    S_{\text{ct}} = \frac{1}{4G_{d+1}} \int_{\partial A} d^{d-1}x^\alpha \sqrt{\tilde{h}}
\end{equation}
where $\tilde{h}$ is the induced metric on the boundary of the entangling region.

\section{Preparing the data}\label{sec:prepare model}

\begin{figure}[ht]
    \centering
    \includegraphics[width=0.7\linewidth]{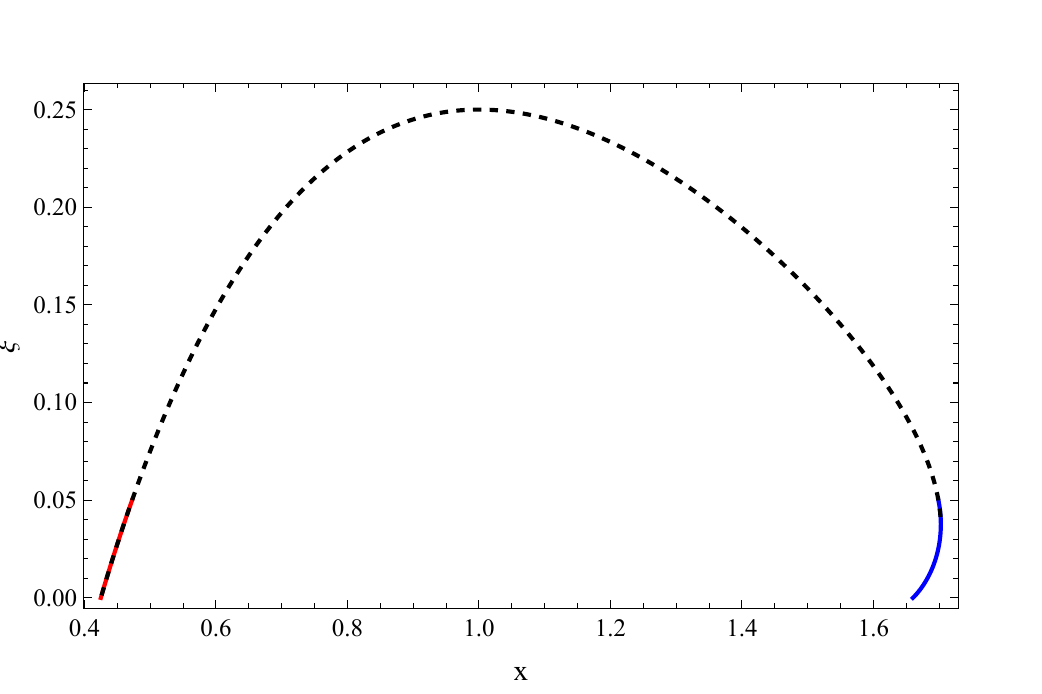}
    \caption{Overlay of the numerical solution (dashed line) with the asymptotic analytical expansion (solid blue/red curves) for both branches, illustrating that there is a match near the boundary. }
    \label{fig:Analytical and numerical data}
\end{figure}
For training data, we will use analytical data obtained from asymptotic analysis near $\xi\sim 0$. 
Numerical data are generated via a Taylor‐expansion algorithm: starting from the prescribed inflection point, both solution branches are constructed (see \cite{landgren2}).  This approach is most accurate in the immediate vicinity of the inflection point. The data develop a second turning point where $(\xi '(x)\to-\infty)$, exactly where the numerical solver breaks down. This divergence occurs close enough to the boundary that the analytical asymptotic expansion remains valid there.  By anchoring our numerics to the analytic solution, we bridge the gap and capture the behavior around this second turning point.

We will work with an inflection point situated at $x_0(\xi_0=\frac{1}{4})= 1$ and endpoints, where $\xi =0$: $x_1=0.424878$ and $x_2=1.660046$. The second turning point is located at $\{x=1.7025, \xi =0.03778\}$.

The boundary conditions we will implement into our loss function are
\begin{align}
    &x_1(\xi=0)=0.424878, \quad x_2(\xi=0)=1.660046 \label{eq: condition 1}\\
    &x(\xi=\frac{1}{4})=1\label{eq: condition 3}\\
    &x(\xi=0.0377816)=1.7025\label{eq: condition 4}\\
    &x'(\xi=0.03778)=0\label{eq: condition 5}.
\end{align}

\begin{figure}[ht]
    \centering
    \begin{subfigure}[b]{0.8\linewidth}
        \centering
        \includegraphics[width=\linewidth]{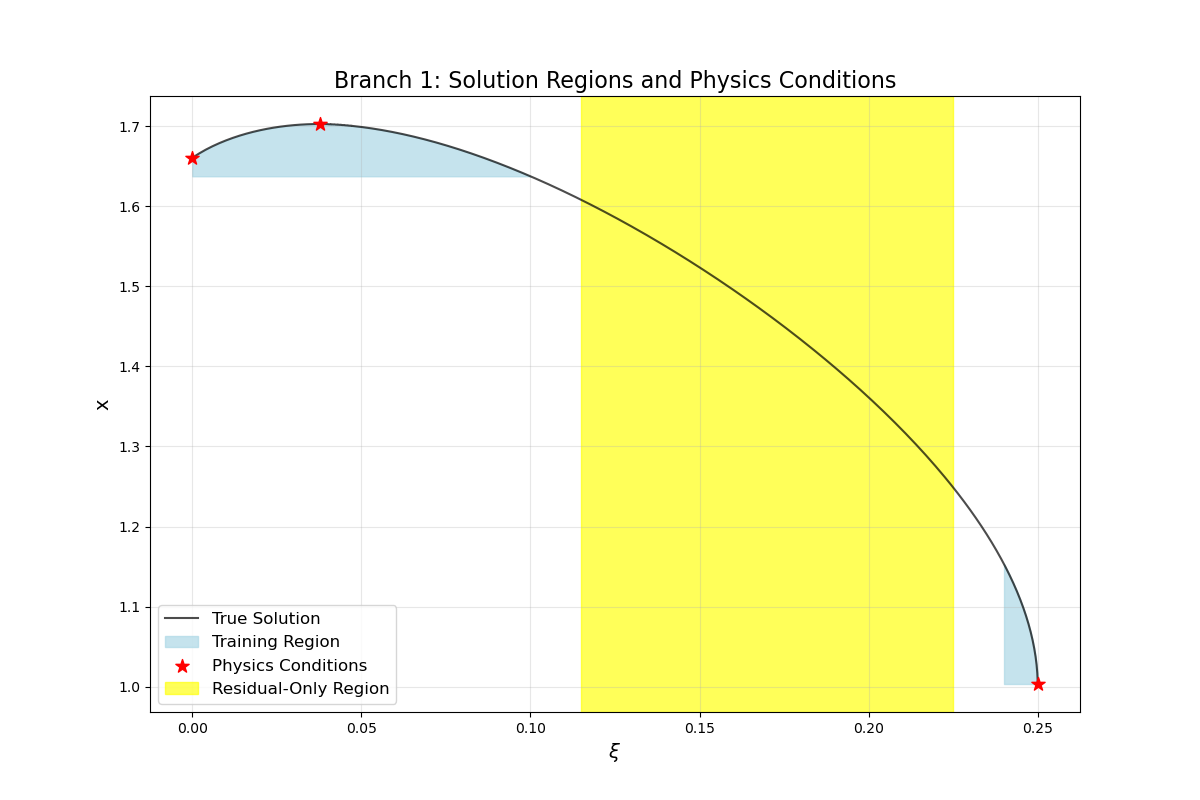}
        \caption{}
        \label{fig:setup a}
    \end{subfigure}
    \vspace{1em}  
    
    \begin{subfigure}[b]{0.8\linewidth}
        \centering
        \includegraphics[width=\linewidth]{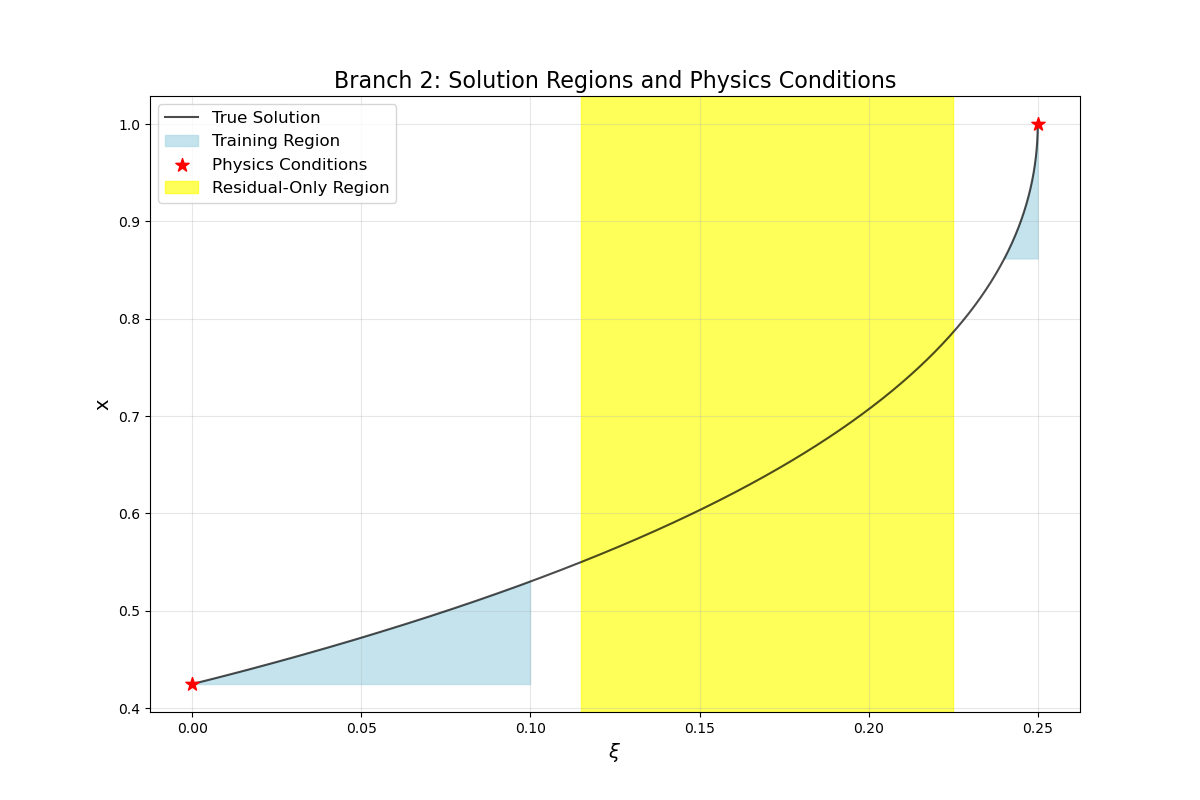}
        \caption{}
        \label{fig:setup b}
    \end{subfigure}
    \caption{(a) Branch 1 and (b) Branch 2: true solution curves $x(\xi)$ with shaded blue regions indicating points used for training data and yellow regions where the PDE residual is enforced in the loss. }
    \label{fig:setup}
\end{figure}

The data training regions, physical collocation points in the loss function as well as the region where the residual is enforces is showed in figure \ref{fig:setup}. In principle, we could enforce the residual everywhere. Our residual weight has been fine tuned to approach zero in the regions rich with training data, whose loss is orders of magnitude smaller than that of the residual.  

We will be working with the Tanh activation function, Adam optimizer \cite{kingma2017adam}, and 2000 epochs around which the mean squared error (MSE) converges. The hyper-parameters and number of residual sampling points in the intermediate regions are fine-tuned and computed over a grid. The two branches will be trained on separately, each with its own network.

\section{B-PINNs}\label{sec:B-PINN}
\begin{figure}[ht]
\centering
\begin{tikzpicture}[
    scale=1,
    every node/.style={transform shape}, 
    neuron/.style={circle, draw, minimum size=1cm, inner sep=0pt},
    arrow/.style={-Stealth, thick},
    label/.style={font=\footnotesize},
    prob/.style={ellipse, draw, minimum size=0.8cm, fill=red!10}
]

\node[neuron, fill=yellow!20] (x) at (0, 0) {$x, t$};
\node[label, above=0.5cm] at (x.north) {Inputs};

\node[neuron, fill=blue!20] (h) at (3, 0) {$\mathcal{N}(\mu, \sigma^2)$};
\node[label, above=0.5cm] at (h.north) {Hidden Layers};

\node[neuron, fill=green!20] (u) at (6, 0) {$\hat{u}(x,t)$};
\node[label, above=0.5cm] at (u.north) {Output};

\draw[arrow, dashed] (h) to[out=30, in=150] (u);
\draw[arrow, dashed] (h) to[out=-30, in=-150] (u);
\node[label, font=\tiny, above=0.1cm] at (4.5, 0.5) {Multiple Passes};

\node[rectangle, draw, minimum height=0.8cm, minimum width=1.5cm, fill=green!20] (avg) at (8, 1) {$E[u]$};
\node[label, font=\tiny, above=0.1cm] at (avg.north) {Ensemble Avg.};

\node[rectangle, draw, rounded corners, minimum height=1.5cm, minimum width=2.5cm, fill=blue!10] (loss) at (9, -1.5) {Physics Loss  $\nabla_\theta L \to 0$};
\node[label, above=0.5cm] at (loss.north) {PDE};

\draw[arrow] (x) -- (h);
\draw[arrow] (h) -- (u); 
\draw[arrow] (u) -- (avg); 
\draw[arrow] (u) -- (loss);

\draw[arrow, dashed] (loss.west) to[out=180, in=-90] (6, -1) to[out=90, in=-90] (h.south);

\end{tikzpicture}
\caption{Schematic of a simple Bayesian Physics-Informed Neural Network (B-PINN). Gaussian-distributed weights ($\mathcal{N}(\mu, \sigma^2)$) enable multiple stochastic forward passes (dashed arrows) which may be used to compute an ensemble average ($E[u]$) for uncertainty quantification, while a physics loss enforces a constraint.}
\label{fig:bayesian network}
\end{figure}
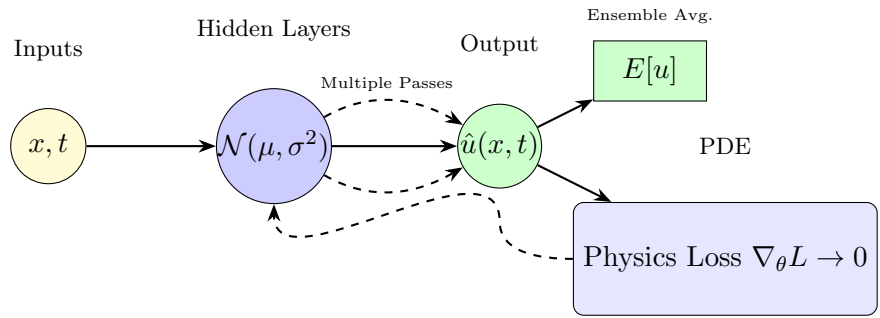

Bayesian neural networks (BNNs), first considered in \cite{Buntine1991BayesianB} introduce a probabilistic approach to modeling by treating the network weights as random variables with specified prior distributions, illustrated in figure \ref{fig:bayesian network}. In short, Bayes' theorem provides a way to calculate the conditional probability of a hypothesis given observed data:
\begin{equation}\label{eq:bayes}
    P(A|B)=\frac{P(B|A)P(A)}{P(B)}.
\end{equation}
The l.h.s. is the posterior probability specifying the uncertainty due to absent or noisy data; the updated belief about $A$ after observing data $B$.  $P(B|A)$ is the likelihood or the probability of observing $B$ given that $A$ is true and specifies the uncertainty owed to noisy data. $P(A)$ is the prior i.e. the initial belief about $A$ before observing $B$ and $P(B)$ is the marginal probability - the total probability of observing $B$, also called the evidence. Bayesian statistics extends Bayes' theorem into a framework to model the probability of an event provided prior knowledge. The prior distributions are updated with observed data and used to form the posterior distributions.

BNNs provide a systematic way to capture the inherent uncertainties and may offer insights into the confidence of the solutions obtained, thereby facilitating more informed decision-making in real world applications \cite{ Linka_2022}.

Furthermore, in the context of B-PINNs, Bayes' theorem (\ref{eq:bayes}) can be expressed as  \cite{arbel2023primerbayesianneuralnetworks}
\begin{equation}\label{eq:B-PINNbayes}
p(\theta|\mathcal{D},\mathcal{P})=\frac{p(\mathcal{D}|\theta)p(\mathcal{P}|\theta)p(\theta)}{p(\mathcal{D},\mathcal{P})}
\end{equation}
where $\theta$ label the weights of the neural network, $\D$ is the training data and $\mathcal{P}$ labels the physical constraints in the loss function. 

The domain on which our solution is supported is given by
\begin{equation}
\Omega=\Omega_{u}+\Omega_{b}+\Omega_{\psi} 
\end{equation}
where $\Omega_{b}$ are the collocation points at the boundaries,  $\Omega_{\psi}$ the collocation points enforced in the loss function not at the boundary and $\Omega_{u}$ the remainder of the training points not subject to constraints in the loss function.
With noisy data, the measurement is taken to have a Gaussian distribution centered around the real value \cite{Yang_2021}: $\bar{u}^i=u(x^i)+\epsilon^i$, where $\epsilon^i$ labels zero-mean independent Gaussian noise, with a standard deviation $\sigma^i$\footnote{we assume that the standard deviation is the same for all subdomains.}. The likelihood in the program is computed as\footnote{since the measurements are taken to be independent the likelihood of the data domain is the product of the likelihood of the subdomains.} \cite{Yang_2021}
\begin{equation}
p(\Omega|\theta)=\prod_{k}p(\Omega_k|\theta), \quad k=u, b, \psi
\end{equation}
where
\begin{equation}
p(\Omega_k|\theta)=\prod_i^{N_k}\frac{1}{2 \pi \sigma_k^i}\exp\left( - \frac{(\hat{u}(x^i)-\bar{u}^i)^2}{2(\sigma_k^i)^2}\right)
\end{equation}
where $N_k$ is the number of points in each subdomain. Weights are learned by maximum likelihood estimation (MLE) \cite{blundell2015weightuncertaintyneuralnetworks}:
\begin{equation}\label{eq:MLE}
\theta^{MLE}=\text{arg} \, \text{max}_{\theta} \log P(\Omega|\theta).
\end{equation}
and the final parameters, $\nu$, of the model are those of a distribution $q(\theta | \nu)$ minimizing the Kullback-Leibler (KL) divergence:
\begin{equation}\label{eq:kl weights}
    \nu^*= \text{arg} \, \text{min}_\nu \text{KL}[q(\theta|\nu)||P(\theta|\Omega)]
\end{equation}
 where
 \begin{equation}
     \text{KL}[q(\theta|\nu)||P(\theta|\Omega)]=\int q(\theta|\nu) \log \frac{q(\theta|\nu)}{P(\theta)P(\Omega|\theta)}d\theta. 
 \end{equation}

To make the weight parameters of our B-PINNs  probabilistic (Gaussian) distributions we use BayesianLinear layers from the blitz-bayesian-pytorch library \cite{esposito2020blitzbdl}, as opposed to e.g. nn.Linear layers typically used for ordinary PINNs. Furthermore our PINN class uses the \textit{@variational\textunderscore estimator} to enable automatic handling of variational inference during training. The loss function is adjusted to include the KL divergence between the approximate posterior and the prior distributions over the weights. The KL divergence acts as a regularization term, penalizing complex models and preventing statistical overfitting, especially important when data is sparse or clustered non-uniformly.

Our training loop performs multiple stochastic forward passes per batch, which approximates the expected loss over the distribution of weights.  Each sample representing a different possible realization of the network weights according to their posterior distributions. A higher number of forward passes leads to a better approximation of the posterior but increases computational cost. The KL divergence term is weighted by a factor $1\times 10^{-6}$ in the case of our entangling surface, to balance its contribution relative to the data fitting and physics-informed components of the loss function. This results in a predictive distribution characterized by a mean and variance, providing a measure of uncertainty in the predictions.

\begin{figure}[h!]
    \centering
    \begin{subfigure}{0.6\textwidth}
        \includegraphics[width=\linewidth]{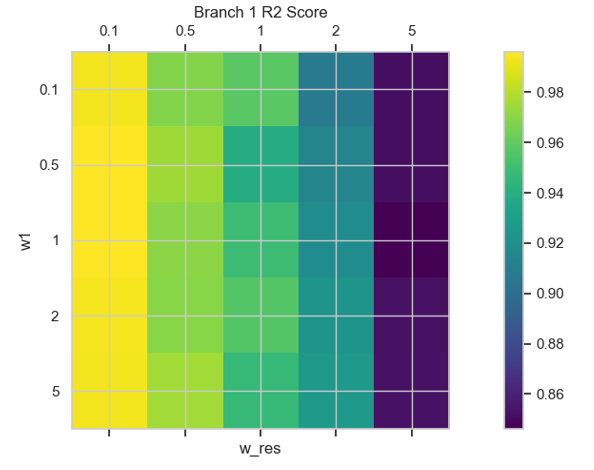} 
        \caption{}
    \end{subfigure}
    \hfill
    \begin{subfigure}{0.6\textwidth}
        \includegraphics[width=\linewidth]{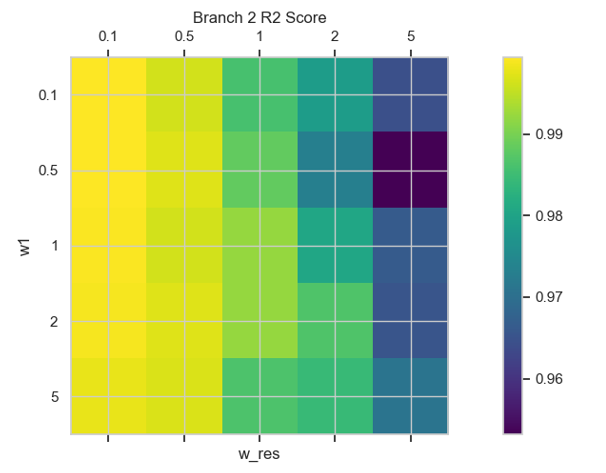} 
        \caption{}
        \label{fig:heatmapbranch2}
    \end{subfigure}
    \caption{$R^2$ score heatmap showing the impact of different values of $w_1$ ($y$-axis) and $w_{res}$ ($x$-axis) on model performance for Branch 1 (top) and Branch 2 (bottom). Lighter colors indicate higher $R^2$ scores, with optimal scores occurring for lower values of $w_{res}$ (0.1 and 0.5) and moderate values of $w_1$ (0.1 to 1). }
    \label{fig:heatmap}
\end{figure}

The learning of the solution to (\ref{xdeqn}) is in particular sensitive to changes in the residual weight, $w_{res}$, whose value dictates how much weight the residual loss contributes to the loss function (and by extension how much weight the model puts on accurately computing the residual). The model is not as sensitive to the relative difference in the weights for condition (\ref{eq: condition 1})-(\ref{eq: condition 4}); in figure \ref{fig:heatmap} they have been put equal to each other. 

Higher values of $w_{res}$ generally result in lower $R^2$ scores, shown by the dark purple shading on the right side. 
Lighter colors (higher $R^2$) are concentrated in the top-left area of the heat map for the first branch, where $w_{res}$ values are lower (0.1 or 0.5) and $w_1$ values are moderate (0.1 to 1). 
Similarly, the second branch shows a similar trend, with the highest $R^2$ scores obtained with lower $w_{res}$ values and moderate $w_1$ values).
The $R^2$ scores for the second branch are generally higher than those for the first one, as indicated by the lighter overall color. The second branch does not have a turning point and is easier to fit.

\begin{figure}[h!]
    \centering
    \begin{subfigure}{0.9\textwidth}
        \includegraphics[width=\linewidth]{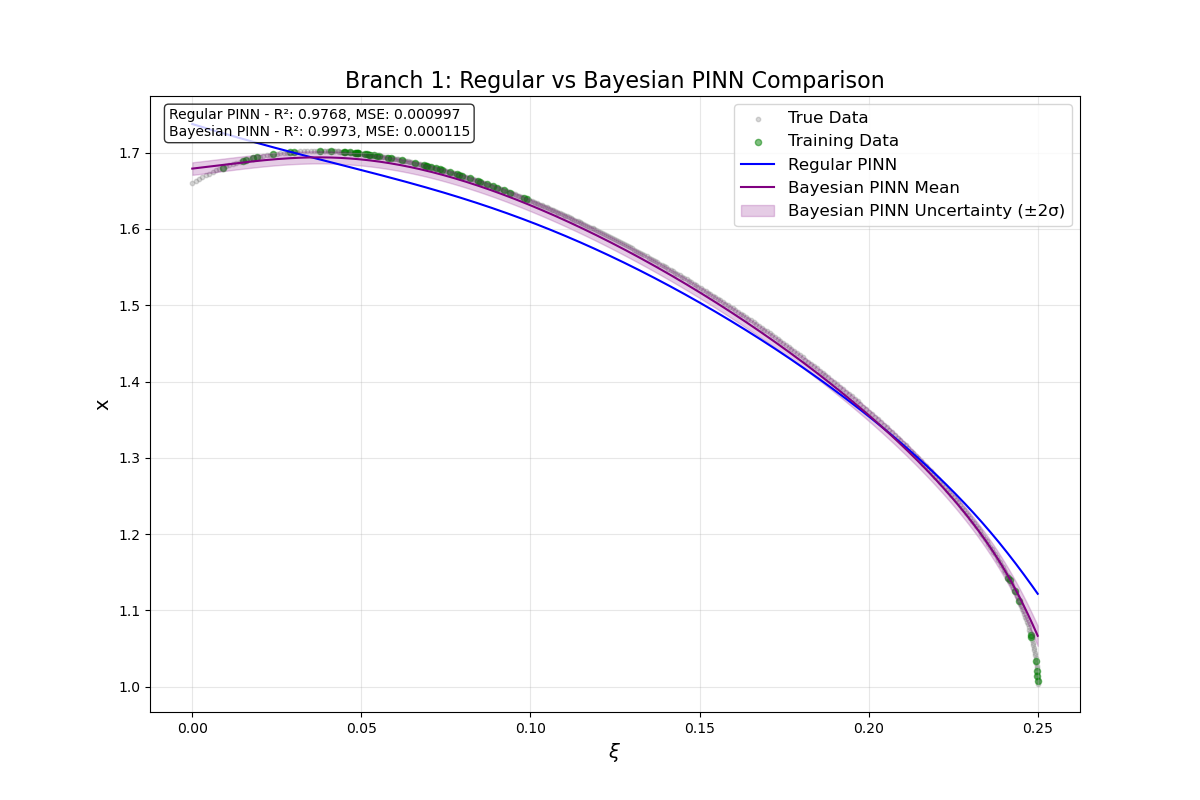} 
        \caption{}
        \label{fig:prediction B-PINN brach1}
    \end{subfigure}
    \hfill
    \begin{subfigure}{0.9\textwidth}
        \includegraphics[width=\linewidth]{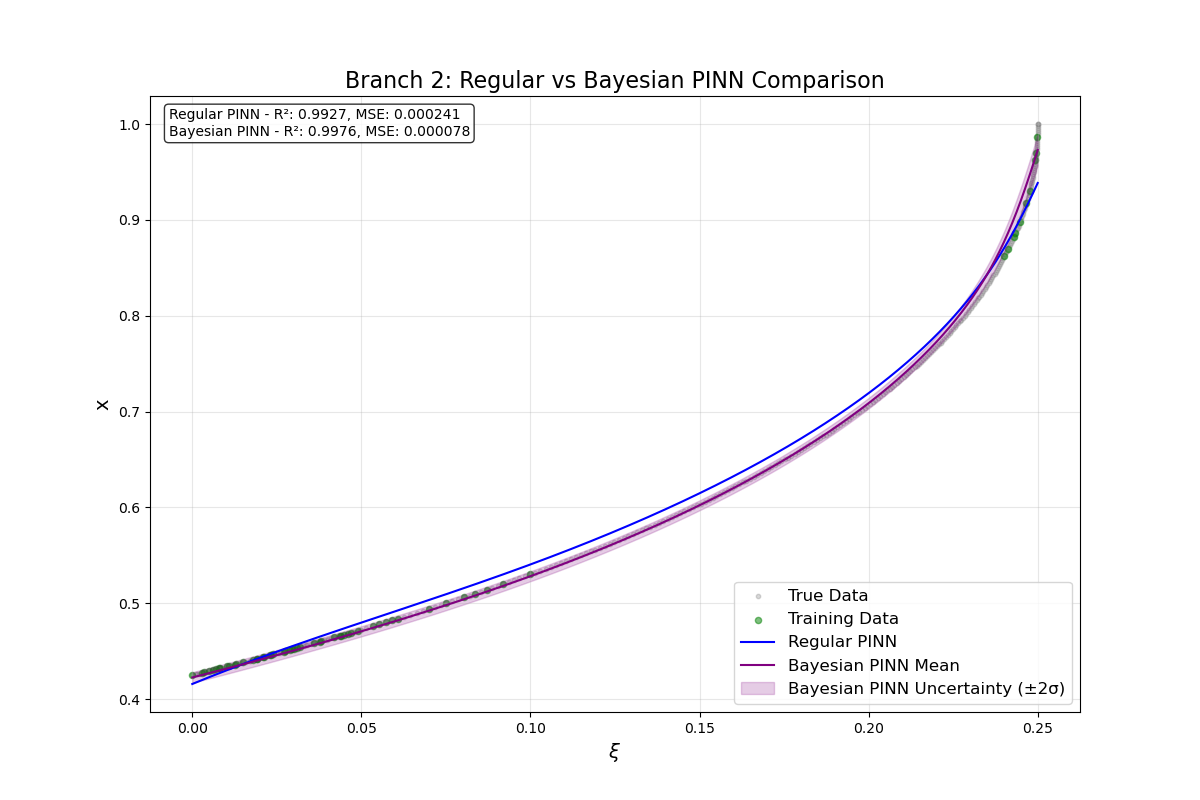} 
        \caption{}
        \label{fig:prediction B-PINN branch 2}
    \end{subfigure}
    \caption{(a) Branch 1 and (b) Branch 2: error between model predictions and true data for regular XPINN (blue) versus Bayesian XPINN (red), highlighting reduced bias of the Bayesian approach. }
    \label{fig:prediction B-PINN}
\end{figure}

\begin{figure}[h!]
    \centering
    \begin{subfigure}{0.66\textwidth}
        \includegraphics[width=\linewidth]{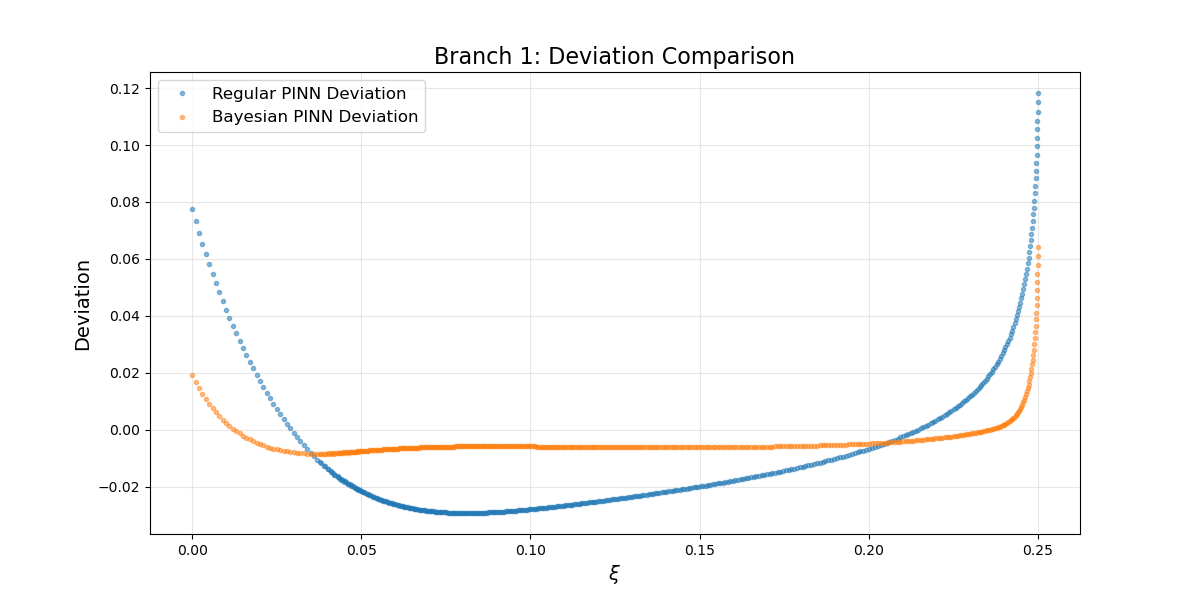} 
        \caption{}
        \label{fig:deviation brach1}
    \end{subfigure}
    \hfill
    \begin{subfigure}{0.66\textwidth}
        \includegraphics[width=\linewidth]{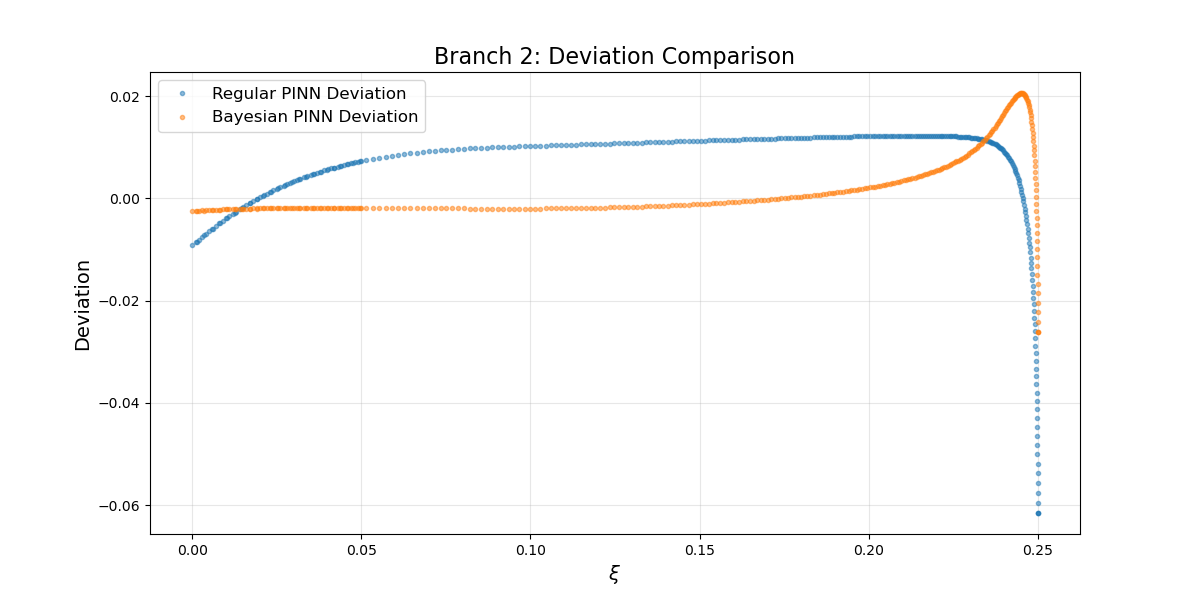} 
        \caption{}
        \label{fig:deviation branch 2}
    \end{subfigure}
    \caption{(a) Branch 1 and (b) Branch 2: plotted residual vs $u$, showing increased residual near steep-gradient regions around the inflection point.}
    \label{fig:deviation}
\end{figure}

\begin{figure}[h!]
    \centering
    \begin{subfigure}{0.66\textwidth}
        \includegraphics[width=\linewidth]{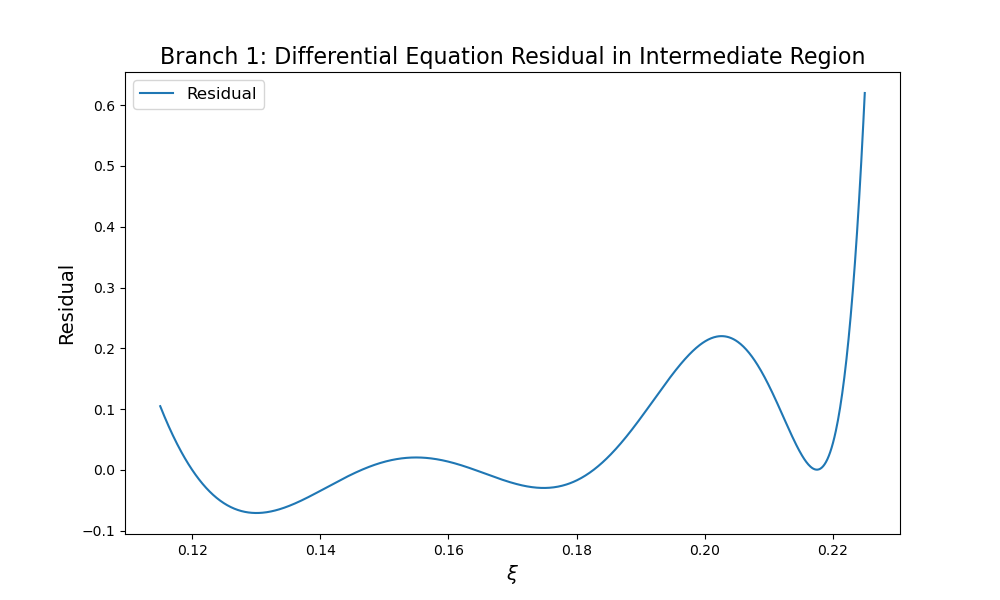} 
        \caption{}
        \label{}
    \end{subfigure}
    \hfill
    \begin{subfigure}{0.66\textwidth}
        \includegraphics[width=\linewidth]{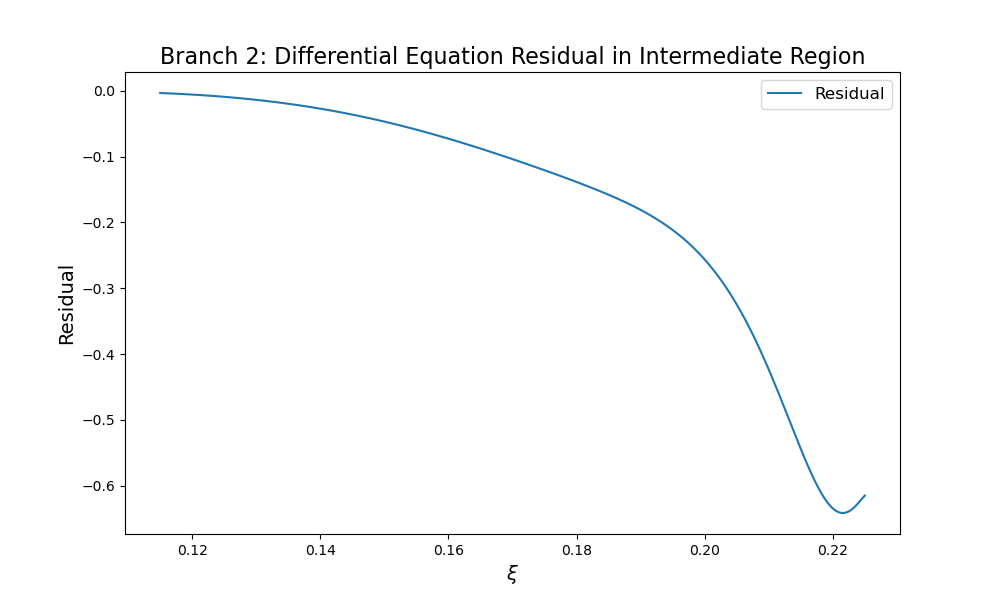} 
        \caption{}
        \label{}
    \end{subfigure}
    \caption{Residual error across the intermediate region for (a) the first branch and (b) the second branch. As expected they increase in regions with steep gradients. }
    \label{fig:residual loss}
\end{figure}

We plot the predicted result in figure \ref{fig:prediction B-PINN} and compare it with a traditional (X)PINN. In figure \ref{fig:deviation} we show the deviation from the true data, and  in figure \ref{fig:residual loss} we display the residual loss in the intermediate regions.  As expected, the deviation increases around the inflection point where the gradients are large.

\section{B-PINNs and confidence}

In purely data-driven machine learning, overconfidence often suggests model misspecification or inadequate uncertainty quantification methods. However, for physics-informed learning, physical knowledge is incorporated into the loss function which can justifiably constrain the solution space so tightly that the posterior distribution collapses around a physically consistent solution. Thus, the model being overconfident by traditional metrics can in some cases be seen as a feature rather than a bug; apparent overconfidence is attributed to the model adhering to the physical constraints.  It was noted in \cite{graf2022errorawareBPINNsimprovinguncertainty} that there are multiple sources of overconfidence in B-PINNs that should no be mixed and an uncertainty quantification framework for Bayesian PINNs that explicitly accounts for the gap between the B-PINN’s prediction and the (unknown) true solution, to mitigate non-justified overconfidence. 

Our approach does not introduce auxiliary error bounds but instead defines a local physical information density and a  physics-constraint coupling (PCC) ratio to diagnose where the model’s existing confidence is driven by its physics constraints versus data, even in complex nonlinear settings where analytical error estimates are unavailable.

The posterior distribution, given data $\mathcal{D}$, and a physics constrain $P$ can be expressed as
\begin{equation}
    p(\theta | \mathcal{D},P) \propto p(\mathcal{D}|\theta)p(\mathcal{P}|\theta) p(\theta)
\end{equation}
and assuming that the prior $p(\theta)$ is a uniform distribution we have 
\begin{equation}
    p(\mathcal{D}|\theta) \propto e^{\left(-L_{\text{ data}}(x)/T \right)}
\end{equation}
where $T$ should be interpreted as some temperature scale or noisy variance, and $L_{\text{data}}$ is the data component of the loss function. Similarly,
\begin{equation}
    p(\mathcal{P}|\theta)\propto e^{-\lambda L_{\text{ODE}}}
\end{equation}
where $L_{\text{ODE}}$ is the physics part of the loss function and $\lambda$ the corresponding weight. 
When the physics constraints are enforced in the learning, the feasible set of parameter configurations, $\theta$, that minimize the terms in $L_{\text{ODE}}$ forms a low-dimensional manifold in parameter space. As $L_{\text{ODE}} \to 0$, the posterior collapses to
\begin{equation}
    p(\theta | \mathcal{D},\mathcal{P}) \propto e^{-(L_{\text{data}}(\theta)+0)/T}
\end{equation}
and the physics effectively prunes the search space of $\theta$, making the posterior sharply concentrated around physically consistent solutions. Near a well-fit solution, $\theta^*$, we have 
\begin{equation}
    \nabla_\theta L_{\text{tot}} |_{\theta^*} \approx 0.
\end{equation}

Using the Hessian, quantifying the local curvature, to assign error bars for a neural network output was first explored in \cite{NIPS1990_7eacb532}. In \cite{10.1162/neco.1992.4.3.448}, the Laplace approximation was implicitly used to obtain a Gaussian centered at the maximum a posteriori (MAP) estimate for a BNN, and below we will use a similar prescription. 

We could expand the loss function into a second-order Taylor series around the MAP estimate $\theta^*$:
\begin{equation}
    L_{\text{tot}}= L(\theta^*) + \frac{1}{2}(\theta-\theta^*)^\perp H(\theta-\theta^*) + \ldots
\end{equation}
where 
\begin{equation}
  H=\nabla_\theta^2 L_{tot}(\theta^*)=\frac{\partial^2 L_{\text{tot}}(\theta)}{\partial \theta \partial \theta^\perp}\vert_{\theta=\theta^*}  
\end{equation}
is the Hessian encoding the local curvature.
Expanding around $\theta^*$ gives us
\begin{equation}
    p(\theta|D,P)=\exp\left( -\frac{1}{T}L_{\text{tot}}(\theta^*) \right)\exp\left( -\frac{1}{2T}(\theta-\theta^*)H(\theta-\theta^*)\right)+ \ldots
\end{equation}
leading to the Laplace approximation
\begin{equation}
    p(\theta|\mathcal{D},\mathcal{P})\approx \mathcal{N} \left(\theta^*, \left( \frac{1}{T} \nabla^2_\theta L_{\text{tot}}(\theta^*)\right)^{-1} \right)=\mathcal{N}\left(\theta^*, \Sigma_\theta \right)
\end{equation}
where $\Sigma_\theta \approx TH^{-1}$.

A strong constraint on the differential equation will increase the curvature as deviations from the true solution rapidly increases $L_{\text{tot}}$ and the Hessian at large $\theta^*$, indicating a sharply peaked posterior.
Predictive variance from the predicted solution, $\hat{u}_\theta(x)$ at any point $x$ is effected by how perturbations in $\theta$ translate into output variations; if the posterior over $\theta$ is highly concentrated, $\hat{u}_\theta(x)$ exhibits low variance. Thus, as the physical constraints are satisfied, the parameter posterior collapses and predictive uncertainty decreases, appearing as overconfidence.

Now, let $f_\theta(x)$ be the neural network's forward pass that approximates $u_\theta(x)$. Linearizing $f_\theta(x)$, around $\theta^*$,  for small $\delta \theta=\theta-\theta^*$ gives   
\begin{equation}
    f_\theta(x) =f|_{\theta^*}(x) + \nabla_\theta f|_{\theta^*}(x)\delta\theta + \ldots.
\end{equation}
where $\nabla_\theta f|_{\theta^*}(x)$ is the gradient of the output with respect to the parameters, evaluated at $\theta^*$. 
The predictive mean can thus be written as 
\begin{equation}
    \mu(x)= E[f_\theta(x)]=f_{\theta^*}(X) + \ldots
\end{equation}
and the predictive variance can be written as 
\begin{align}
\label{eq: sigma squared}
    \sigma^2(x)&=\text{Var}[f_\theta(x)]\\
    &=\nabla_\theta f|_{\theta^*}(x)^\perp (TH^{-1})\nabla_\theta f|_{\theta^*}(x) + \ldots.
\end{align}
Hence, larger curvature in $L_{\text{tot}}$ (i.e. larger $H$) leads to smaller variance in $\sigma^2(x)$. In other words, strong physics constraints force the models posterior to collapse around a solution satisfying the differential equation and boundary condition. 

To illustrate how the residual link to the parameter space curvature, we may consider a generic PDE operator\footnote{assuming that the residual is enforced at the boundaries as well (which is not the case in our entangling surface example).}

\begin{equation}
    R(x, \hat{u}_\theta(x))=\mathcal{N}(\hat{u}_{\theta}(x))
\end{equation}
where we enforce $R(x, \hat{u}_\theta(x))=0$ for $x \in \Omega$. We might write the PDE and boundary terms in the loss functions as

\begin{equation}
    L_{\text{PDE}}(\theta) = \int_\Omega \left( R(x, u(x)) \right) dx, \quad L_{\text{BC}}=\sum_i \left( R_{\text{BC} }(u(x_i))\right)^2.
\end{equation}
Taking the gradient w.r.t. $\theta$ gives
\begin{equation}
    \nabla_\theta L_{\text{PDE}}(\theta) = \int_\Omega 2R(x, u(x))\nabla_\theta R(x, u(x))dx
\end{equation}
where
\begin{equation}
    \nabla_{\theta}R(x, u(x))=\frac{\partial R}{\partial u} \nabla_{\theta}u(x).
\end{equation}
Strong PDE constraints imply that $||\partial R / \partial u||$ is large near a valid solution, thus inflating the Hessian $\nabla_\theta^2L_{\text{tot}}$ driving a sharply peaked posterior. Note that this analysis considers only the direct dependence on $u$; for PDEs with higher-order derivatives, one might also consider terms like $\partial R / \partial u'$, $\partial R / \partial u''$, etc., which can also contribute to the Hessian's structure. 

To diagnose the relationship between physical fidelity and predictive certainty, we may define a physical information density, that takes all physics constraints  into account as 
\begin{equation}\label{eq:sensitivity}
I(x)\equiv \sum_i||\nabla_{\hat{u}}^i\chi^i(\hat{u}_\theta(x))||^2
\end{equation}
where $\chi^i$ is any local operator enforcing a physical constraint over some $x$ (this could for instance be the differential or boundary operator). The l.h.s gauges the sensitivity of the physical constraints to perturbations in $u$, and should remain large as the solution aligns more closely with the physical conditions.

We may think of $I(x)$ as indicating how stiff the physics conditions are at point $x$. When $I(x)$ is high, even a tiny deviation in the solution $u(x)$ significantly increases the loss of the physical conditions, leaving little room for variation.

The epistemic predictive variance $\sigma^2(x)$ reflects how uncertain the model is about its prediction at a point $x$. In other words, if $I(x)$ is large, then any deviation $\delta u$ impose a large Hessian. As a consequence, small parameter perturbations, $\delta \theta$ that would significantly change the predicted solution at points of high $I(x)$ are penalized. 

A strong local constraints (high $I(x)$) lead to a sharply peaks posterior and lower variance, reflecting a local curvature effect near the solution manifold.

However,  a high $I(x)$ does not guarantee low uncertainty. In regions where physics is complex, such as near sharp or fluctuating gradients, both $I(x)$ and $\sigma(x)^2$ can be large.

This complexity increases the network's sensitivity to parameter changes, increasing $\sigma(x)^2$. Thus, while $I(x)$  measures the stiffness of the physics, the predictive variance depends on the interplay between the curvature $H$  and the output's sensitivity to parameters, $\nabla_\theta f|_{\theta^*}(x)$, as evident in the variance expression. However, it is important to note that even if the uncertainty and physical stiffness are high in the same regions, uncertainty would be even higher without physical constraints. We will comment more on this in section \ref{sec:probing overconfidence}.

To \textbf{diagnose} the overall confidence and whether or not it is due to external constraints on the loss functions,  we may define a global physics-constraint coupling (PCC):

\begin{equation}\label{eq:pcc gen}
    \text{PCC}_\Omega\equiv \frac{\int_\Omega I(x)dx}{\int_{\Omega}\sigma^2(x)dx}
\end{equation}

where a higher PCC suggests that the model's overconfidence can be driven by strong physical constraints rather than by data abundance or calibration artifacts. Furthermore, the governing equations and conditions have tightly constrained the solution space, leaving little flexibility for variation.  In particular, high confidence in regions with low information density may signal overconfidence and should be treated with caution, whereas high confidence in regions with rich information content is more likely to be justified and expected.

It is important to note that different PDEs may benefit from alternative definitions of the information density, as the specific structure of the differential operators can vary significantly between problems. For instance, in the case of the Van der Pol equation (\ref{eq:van der pol}), the functional derivative of the residual with respect to the output, $u$, is constant. For such equations, one may obtain richer insights by considering constraints beyond the residual alone. For other PDEs, a more informative definition may include derivatives with respect to higher-order terms: 
\begin{equation}
I(x) = \sum_{k \in D_i} \left\rvert\left\rvert \frac{\partial \chi^i}{\partial u^{(k)}}\right\rvert \right\rvert^2,
\end{equation}
where $D_i$ is the set of derivative orders that operator $i$ depends on. However, applying this particular definition to some PDEs, such as the Van der Pol equation, would cause the residual contributions to dominate the boundary conditions, obscuring their effect. 

The choice of definition should be guided by the specific structure of the PDE. If we can demonstrate that epistemic uncertainty is low in regions where physics conditions are present and, in particular, where these conditions have a strong impact on the solution manifold, then apparent overconfidence in such regions can be expected. The appropriate method to probe the strength of a physics condition's impact may vary from equation to equation. The information density is not intended to provide a quantitatively precise ranking of how individual constraints compete in shaping the solution manifold. Rather, it serves as a diagnostic tool to identify where the physics most strongly influences the posterior distribution.

\subsection{Probing overconfidence}\label{sec:probing overconfidence}
To further understand apparent over-confidence, we may look at more calibration metrics. 

For our B-XPINN the validation set is $\{ x_i, u_i \}_{i=1}^N$, and via our ensemble sampling, obtained with $M$ stochastic forward passes during learning, we have  
\begin{equation}
    \hat{u}_{i,1}, \hat{u}_{i,2}, \ldots, \hat{u}_{i,M}.
\end{equation}
The predictive mean is 
\begin{equation}
    \mu(x_i)\approx \frac{1}{M}\sum_{j=1}^Mu_{i,j}
\end{equation}
and the predictive standard deviation is
\begin{equation}
    \sigma(x_i) \approx \sqrt{\frac{1}{M} \sum_{j=1}^M(\hat{u}_{i,j}-\mu(x_i))^2}.
\end{equation}
Consider a probabilistic model that, for each input $x_i$, provides the predictive distribution $p(\hat{u}|x_i)$. In a Bayesian or ensemble-based neural network, this distribution often take the form of a Gaussian approximation $\mathcal{N}(\mu (x_i), \sigma^2(x_i))$, or a collection of samples from which one can estimate prediction intervals. A coverage or quintile-based definition of calibration examines how well the predicted intervals match the empirical frequency with which that true target fall into those intervals. 

Defining a nominal coverage level $\alpha \in [0,1]$ (see e.g. \cite{arie2024confidenceintervalssimultaneousconfidence} for a discussion on coverage intervals and useful uncertainty in deep learning), with $\alpha=0.9$ corresponding to a $90$ percent prediction interval, the $\alpha$-coverage interval for each data point $x_i$ is
\begin{equation}
    I_\alpha (x_i)= [\mu(x_i)-z_\alpha \sigma(x_i), \mu(x_i)+ z_\alpha \sigma(x_i)],
\end{equation}
where $z_\alpha$ is the quantile factor (e.g. $z_{0.9}\approx 1.645$ for a one-sided Gaussian). More generally, if the model is assumed to be Gaussian, one can directly compute the lower and upper $\alpha$-quantiles from the predictive samples. The observed coverage is the fraction of data points whose true values $u_i$ lies within the $\alpha$-coverage interval:
\begin{equation}
    \text{ObservedFrequancy}(\alpha) =\frac{1}{N}\sum_{i=1}^N\{u_i \in I_\alpha(x_i) \}
\end{equation}
where $N$ is the number of data points considered.
A model is said to be perfectly calibrated if $ \text{ObservedFrequancy}(\alpha)=\alpha, \quad \forall \alpha\in[0,1]$. In practice we visualize this in a calibration plot (sometimes called reliability diagram), which plots $\text{ObservedFrequancy}(\alpha)$ against $\alpha$.
If the curve lies below the diagonal line, the intervals are too narrow, indicating overconfidence. If the curve lies below the diagonal line, the intervals are too narrow, indicating overconfidence. If it lies above the diagonal, the intervals are too wide, indicating underconfidence. 

\begin{figure}[h!]
    \centering
    \begin{subfigure}{0.45\textwidth}
        \includegraphics[width=\linewidth]{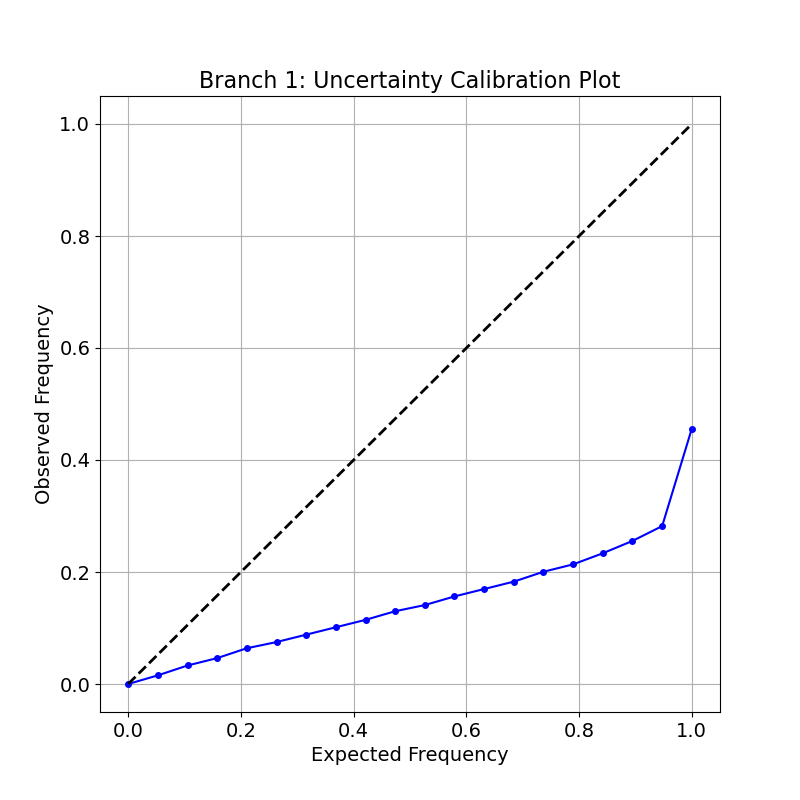} 
        \caption{Uncertainty calibration plot for the first branch, showing consistently overconfidence.}
        \label{fig:calibration plot 1}
    \end{subfigure}
    \hfill
    \begin{subfigure}{0.45\textwidth}
        \includegraphics[width=\linewidth]{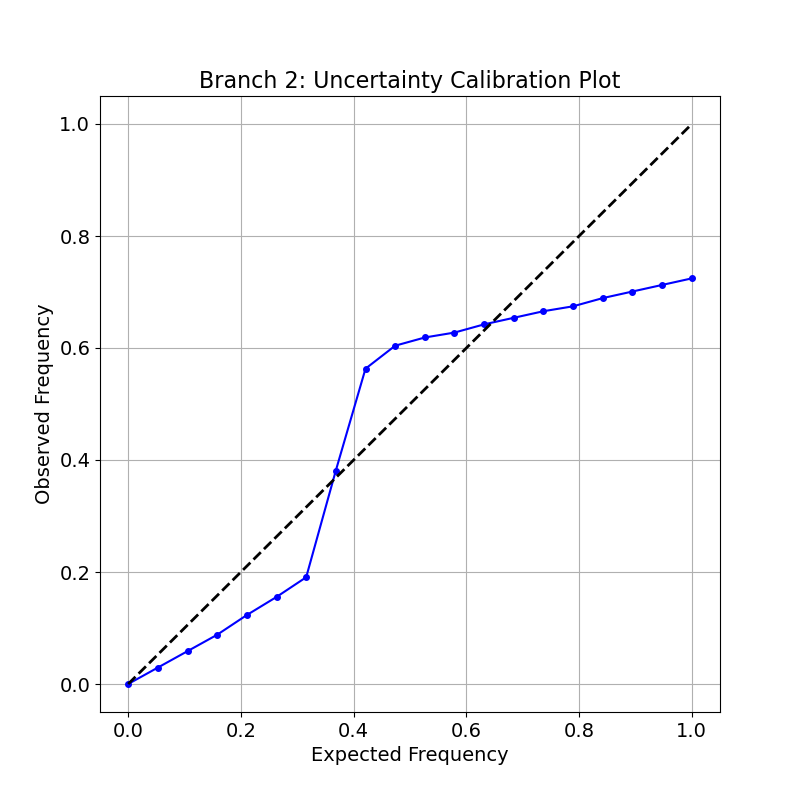} 
        \caption{Uncertainty calibration plot for the second branch, indicating mostly overconfidence with an oscillation around $\alpha=0.45$.}
        \label{fig:calibration branch 2}
    \end{subfigure}
    \caption{}
    \label{fig:calibration plots}
\end{figure}

For the first branch of the solution of the entangling surface, we see in figure \ref{fig:calibration plot 1} that the calibration curve is consistently below the diagonal line, indicating strong and consistent overconfidence, while for the second branch, in figure \ref{fig:calibration branch 2} shows a mostly overconfident behavior, except in a small region near $\alpha=0.45$. The latter is not unexpected as the second branch has fewer physics conditions that the first branch (recall that the second branch has two conditions at the second turning point). 

\begin{figure}[h!]
    \centering
    \begin{subfigure}{0.6\textwidth}
        \includegraphics[width=\linewidth]{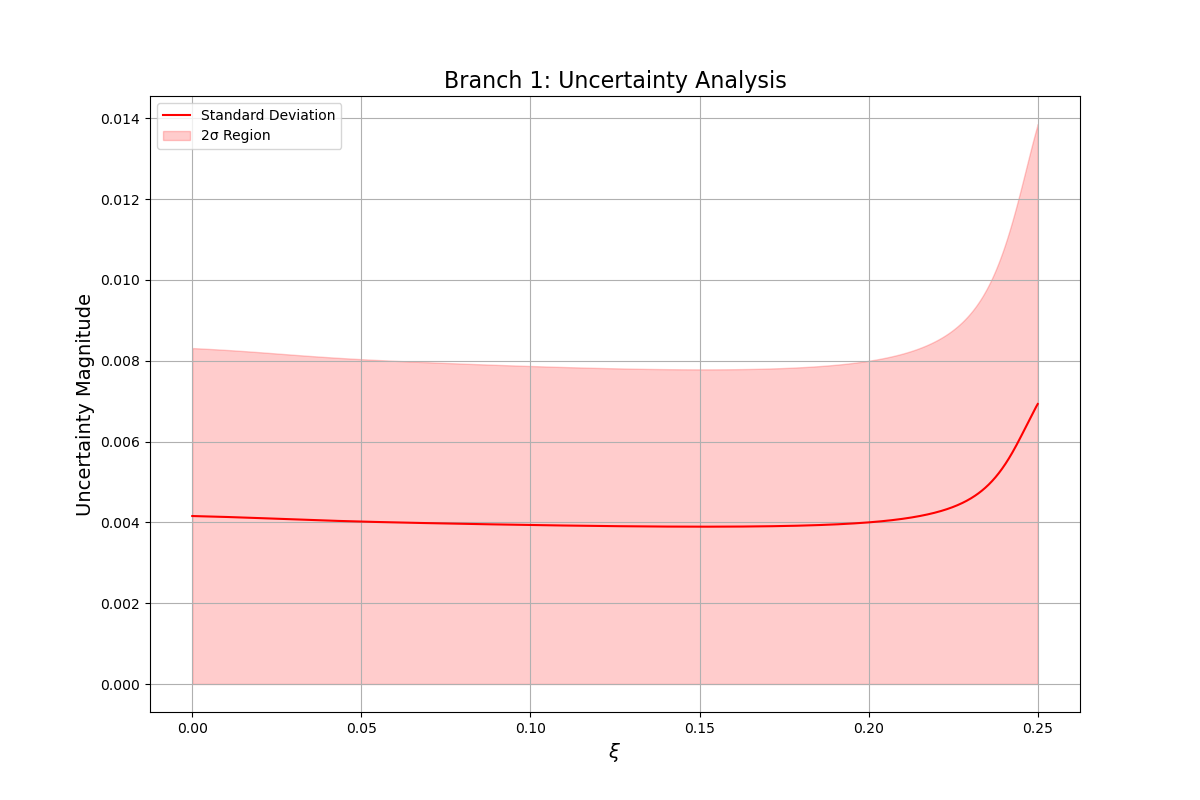} 
        \caption{}
        \label{}
    \end{subfigure}
    \hfill
    \begin{subfigure}{0.6\textwidth}
        \includegraphics[width=\linewidth]{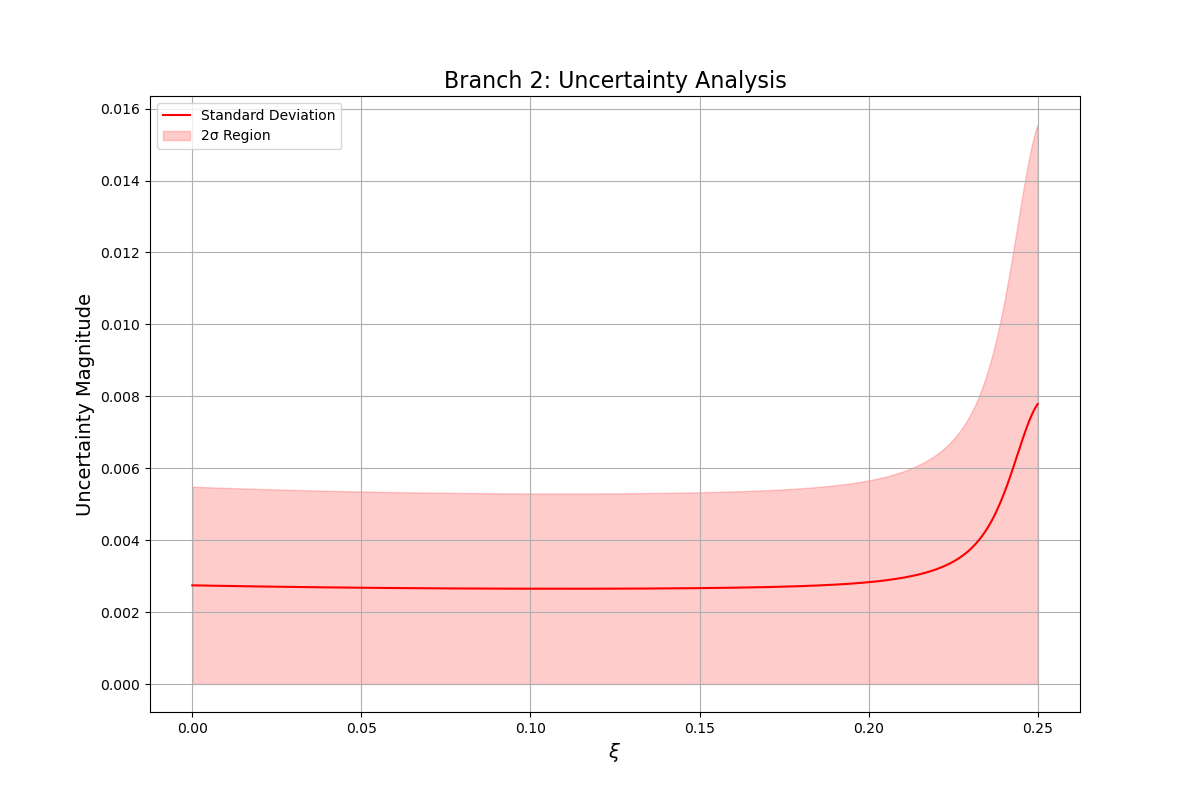} 
        \caption{}
        \label{}
    \end{subfigure}
    \caption{(a) Branch 1 and (b) Branch 2: shaded uncertainty bands, exhibiting wider uncertainty near boundary regions with complex ODE behavior.}
    \label{fig:uncertainty}
\end{figure}

For the entangling surface discussed in the previous section, we have tuned the hyperparameters to optimize the $R^2$ scores, while maintaining the highest prediction interval coverage probability. The latter defines the fraction of validation data points for which the true value falls within the predicted confidence interval prediction interval \cite{kuleshov2018accurateuncertaintiesdeeplearning}.  The confidence band is plotted in figure \ref{fig:uncertainty} and  despite their narrow width, they follow a pattern that makes physical sense, with increased uncertainty near the boundaries,  regions of complex ODE behavior. 

\begin{figure}[ht]
    \centering
    \begin{subfigure}{0.7\textwidth}
        \includegraphics[width=\linewidth]{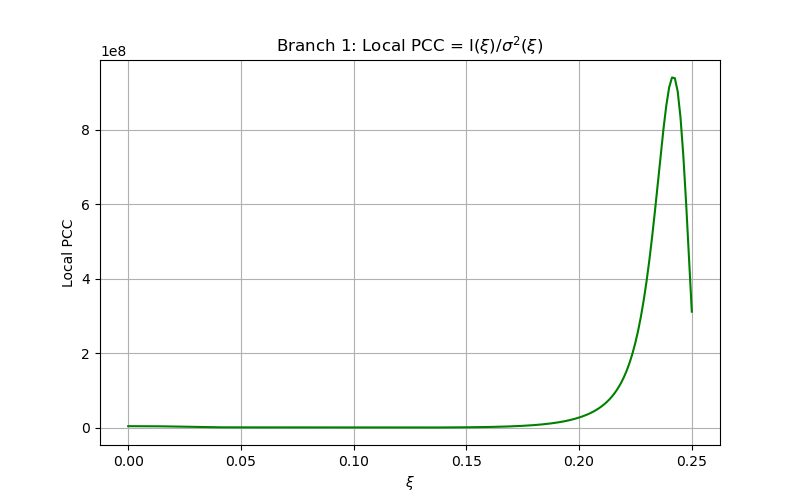} 
        \caption{}
        \label{fig:local pcc}
    \end{subfigure}
    \hfill
    \begin{subfigure}{0.7\textwidth}
        \includegraphics[width=\linewidth]{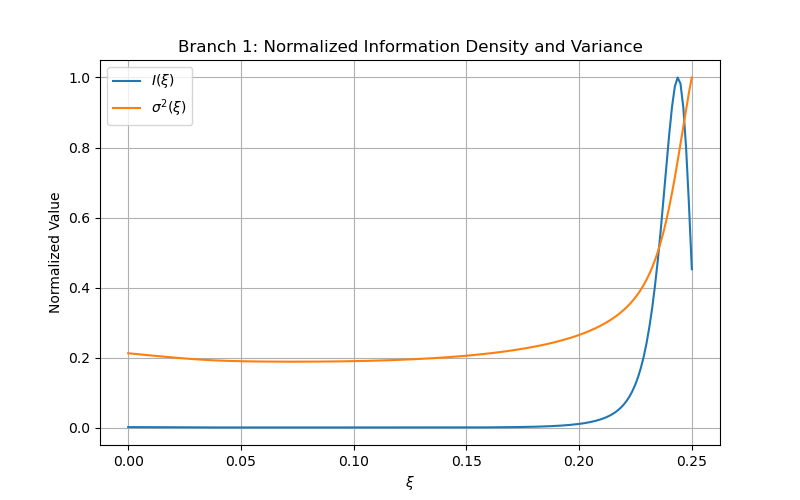} 
        \caption{}
        \label{fig:I and sigma vs u}
    \end{subfigure}
    \caption{(a) PCC$(\xi)$ vs $\xi$ for Branch 1: physics-constraint coupling grows about the inflection point, and decreases as we switch from residual constraints to point-wise conditions. (b) Normalized information density $I(u)$ (blue) and predictive variance (yellow) for Branch 1, increasing monotonically. }
    \label{fig:pcc metrics e surface}
\end{figure}

In figure \ref{fig:local pcc} we still observe that the local PCC reaches its highest value about the inflection point, where we have three boundary conditions clustered. This peak confirms that, in that narrow region, the enforced physics constraints collapse the posterior most strongly. In figure \ref{fig:I and sigma vs u} we plot the normalized predictive variance $\sigma^2(u)$ against the normalized information density $I(u)$\footnote{$I(x)$ is many orders of magnitude larger than $\sigma(x)^2$ and to appropriately compare them, we deploy a simple max-based normalization: $I(x) \to \frac{I(x)}{\text{max}[I(x)] + \varepsilon}$. } \footnote{The $u$ here should not be confused with the predicted output, which in the case of the entangling surface is the independent variable in the differential equation. }. The $I(u)$ profile is very small up to $u\approx 0.2$ after which it climbs sharply as the residual constrains begin to carve out the solution manifold, before slightly dipping in the band $0.24\lesssim
u \lesssim 0.25$ where the loss switches from a distributed residual to a point wise boundary‐condition enforcement. In contrast $\sigma^2(u)$ grows monotonically towards its maximum at the inflection point. The dip in information density and local PCC about the inflection point does not necessarily mean that physics constraints are weaker at the boundary point, but simply that a point wise constraint contributes less to the gradient‐based stiffness than the residual constraints.

\subsection{Further examples}\label{sec: further examples}

\subsubsection{Liouville-type equation}
We expand on the general analysis above by considering a simpler non-linear Liouville-type differential equation, given by
\begin{equation}
    u''(x)+Ke^{u(x)}=0, \quad x\in[0,1]
\end{equation}
with $K=1$ and boundary conditions u(0)=0, u(1)=0. While this equation does not have a simple closed-form solution, one can easily obtain a true numerical solution for reference. 
In this simple example we have
\begin{equation}
    \mathcal{N}(u)=u''(x)+e^{u(x)}, \quad \partial_u \mathcal{N}(u)=e^{u(x)},
\end{equation}
and the information density yields
\begin{equation}
    I(x)=e^{2u(x)}.
\end{equation}

\begin{figure}
    \centering
    \includegraphics[width=0.6\linewidth]{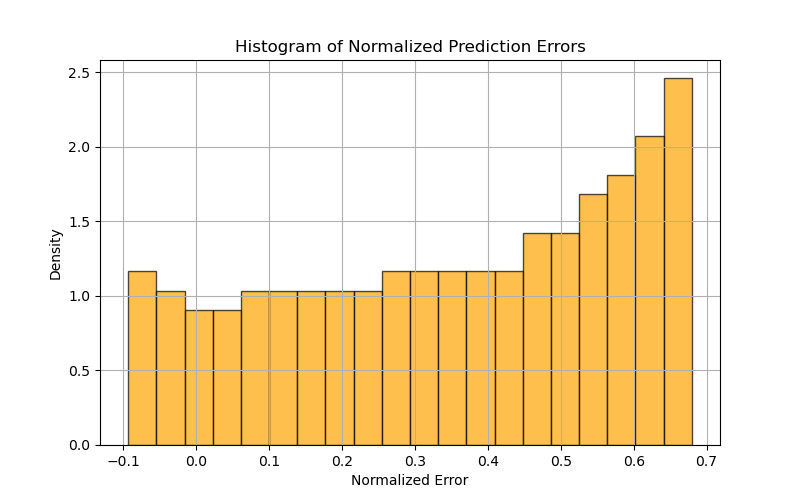}
    \caption{Normalized error distribution for the Liouville-type equation, revealing a skewed, heavy-tailed distribution indicative of systematic under-estimation of uncertainty (overconfidence).}
    \label{fig:histogram}
\end{figure}

In figure \ref{fig:histogram} we display the histogram of normalized prediction error: $\frac{1}{\sigma(x)}(\hat{u}(x)-u(x))$. If the model’s predictive uncertainties are well-calibrated (i.e., the predicted standard deviations truly reflect the variability and confidence levels), we would expect a bell-shaped histogram centered at zero, resembling a Gaussian distribution. However, the observed heavy concentration of negative normalized errors indicates a systematic bias and an underestimation of uncertainty. This suggests that the model’s posterior is excessively narrow i.e., a sign of overconfidence.

\begin{figure}
    \centering
    \includegraphics[width=0.66\linewidth]{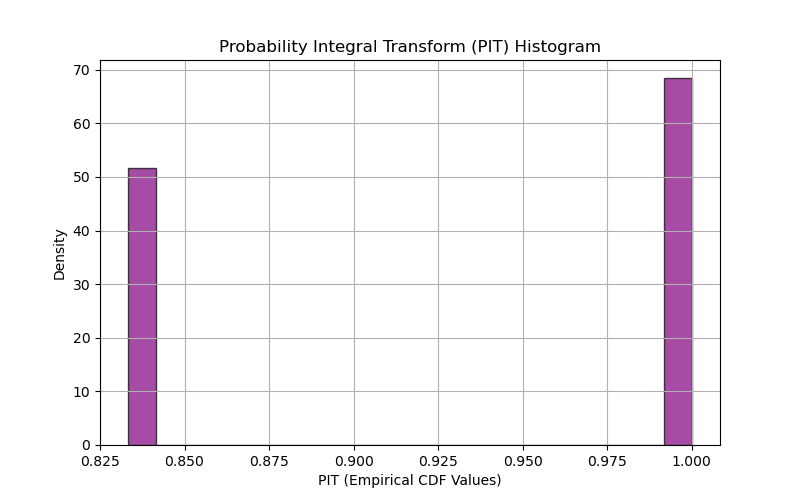}
    \caption{Distribution of PIT values for the Liouville example, displaying two sharp peaks rather than uniformity, confirming miscalibration.}
    \label{pit histogram}
\end{figure}

The probability integral transform (PIT) histogram is another diagnostic for calibration. For each test point, $x$, the model produces a predictive distribution $p(u|x)$ with cumulative distribution function (CDF) $F(u|x)$. For the true observed value, the PIT value is defined as 
\begin{equation}
    p_{\text{PIT}}(x) = F(u_{\text{true}}|x).
\end{equation}
For a Gaussian predictive function, $ p(u|x)=\mathcal{N}\left( \hat{u}(x), \sigma(x)^2 \right)$,
with the corresponding CDF:
\begin{equation}
    F(u|x)= \Phi\left(\frac{u-\hat{u}(x)}{\sigma(x)} \right),
\end{equation}
where $\Phi$ is the CDF of the standard normal distribution. For a given test point, the PIT value thus yields
\begin{equation}
    p_{\text{PIT}}(x)=\Phi \left(\frac{u_{\text{true}}(x)-\hat{u}(x)}{\sigma(x)} \right).
\end{equation}
A well calibrated statistical model if $p_{\text{PIT}}(x)$ is uniformly distributed over $[0,1]$.
In figure \ref{pit histogram} would thus be expected to be flat. However, the distinct peaks strongly indicates that the predictions are miscalibrated.

\begin{figure}
    \centering
    \includegraphics[width=0.5\linewidth]{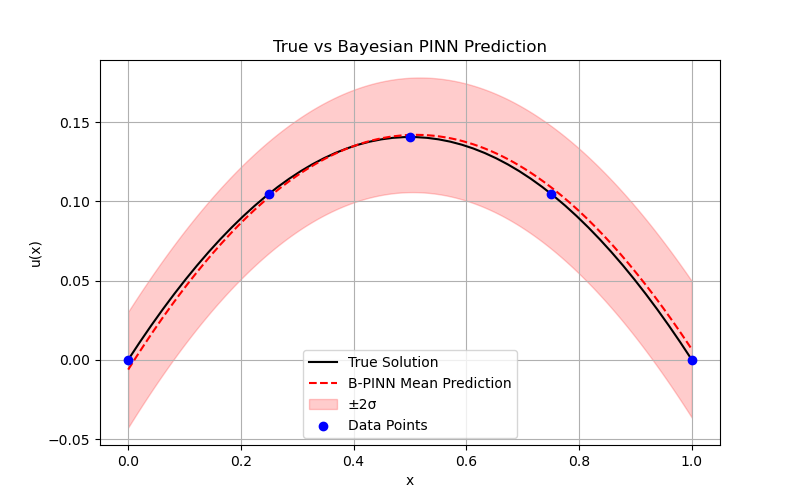}
    \caption{Predicted solution vs true solution, with the uncertainty band.}
    \label{fig:non-lin true}
\end{figure}

As can be seen in figure \ref{fig:non-lin true}, the model performs well and the network resembles the true solution.  Here we get a high global PCC of order $\mathcal{O}(10^3)$, and in this simpler example we have similarly demonstrated that the solution is heavily constrained by the physics, with an overconfident posterior distribution and the model’s confidence grows as it more strictly adheres to the physical laws. 

\begin{figure}
    \centering
    \includegraphics[width=0.5\linewidth]{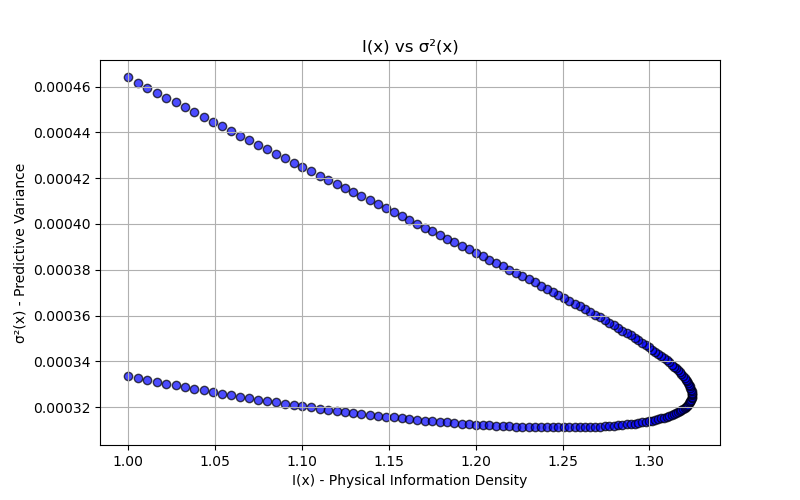}
    \caption{Parametric plot of information density vs uncertainty for the Liouville type equation, illustrating a non-monotonic inverse trend punctuated by complex behavior..}
    \label{fig:parametric plot non linear}
\end{figure}

Figure \ref{fig:parametric plot non linear} displays a parametric plot: $x \to \{I(x), \sigma^2(x) \}$, showing a non-monotonic and non-linear relationship between the information density and uncertainty; hence the turning point behavior.

\subsubsection{Van der Pol equation}
As a next example, we consider a single period of the Van der Pol equation, which exhibits more complex behavior than a simple harmonic oscillator. This equation is widely used to model nonlinear dynamical systems in various fields, including biology (e.g., cardiac rhythms) and electronics (e.g., vacuum tube circuits) \cite{chughtai2023examiningvanderpol}.

Over a cycle, the solution exhibits structural features reminiscent of the entangling surface discussed in section \ref{sec: entangling surface}, particularly in terms of broken symmetry around turning or inflection points. 

The Van der Pol equation is given by: 
\begin{equation} \label{eq:van der pol}
u'' - \mu (1 - u^2) u' + u = 0 
\end{equation} 
where $\mu$ controls the strength of nonlinearity.

The equation is sufficiently non-trivial while remaining analytically and numerically well understood. Moreover, high-quality numerical solutions can be readily obtained using standard ODE solvers. Its solution contains regions of varying dynamical behavior, naturally leading to variations in the information density 
$I(t)$. 

\begin{figure}
    \centering
    \includegraphics[width=0.6\linewidth]{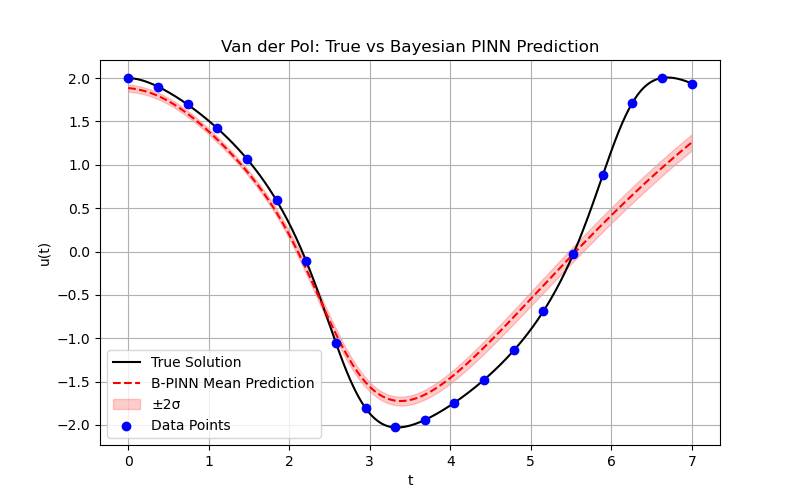}
    \caption{True vs predicted Van der Pol solution. with boundary conditions only around $t=0:  u(0)=2, u'(0)=0$. }
    \label{fig: true vs predicted van der Pol}
\end{figure}

\begin{figure}
    \centering
    \begin{subfigure}{0.66\linewidth}
        \centering
        \includegraphics[width=\linewidth]{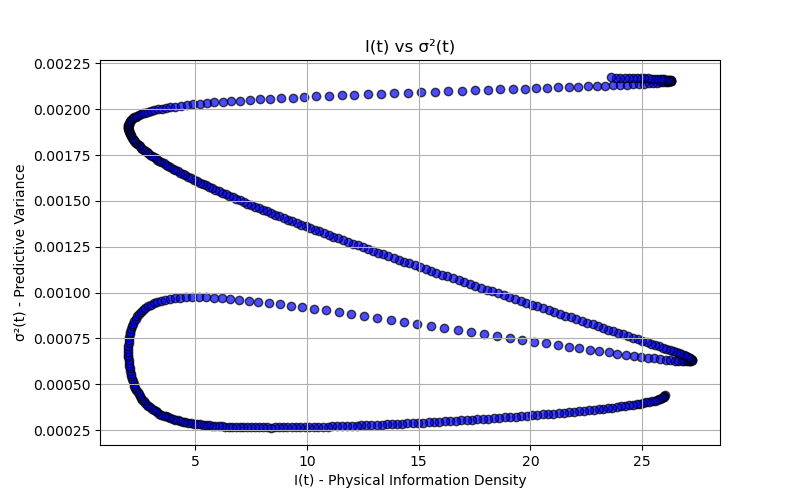}
        \caption{Parametric plot of $I(x)$ and $\sigma^2$, showing non-linear coupling across dynamic regimes..}
        \label{fig: parametric van der Pol}
    \end{subfigure}
    \begin{subfigure}{0.66\linewidth}
        \centering
        \includegraphics[width=\linewidth]{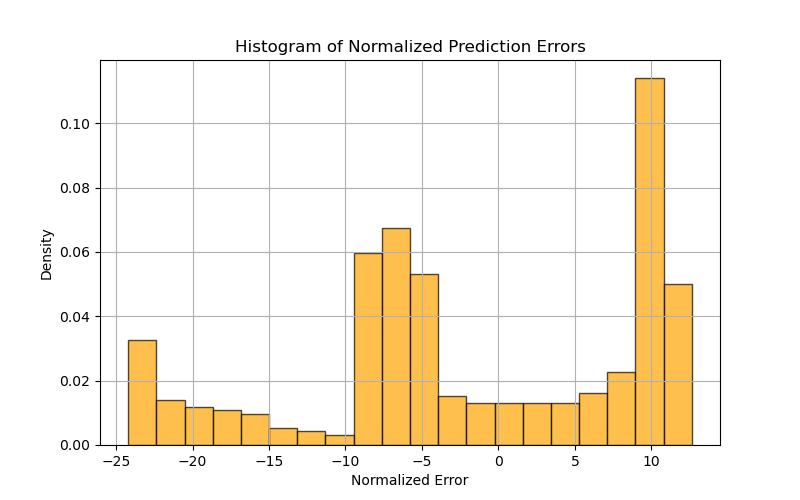}
        \caption{Distribution of normalized prediction errors, again showing signs of an overconfident model.}
        \label{fig: histogram van der Pol}
    \end{subfigure}
    \caption{}
    \label{fig:van der Pol}
\end{figure}

We prepare the data and use the initial conditions $u(0)=2$, $u'(0)=0$. Similar to the previous example, the histogram in figure \ref{fig: histogram van der Pol} shows that the model is statistically overconfident, and the parametric plot in figure \ref{fig: parametric van der Pol} shows that while higher $I(t)$ often corresponds to lower uncertainty, the dynamics of the Van der Pol equation introduce regimes where the relationship between physical constraints and predicted variance is more complex. Similarly to the previous example, we do not see a straightforward inverse relationship between uncertainty and $I(t)$  in figure \ref{fig: parametric van der Pol}. In figure \ref{fig: true vs predicted van der Pol} we see the true numerical solution vs the predicted solution. Although the collocation points for the residual are enforced throughout the domain, we have only enforced initial conditions around $t=0$ and not around the boundary $t=7$; it is evident that the accuracy quickly can deteriorate when physics conditions are absent.

\begin{figure}[ht]
    \centering
    \begin{subfigure}{0.7\textwidth}
        \includegraphics[width=\linewidth]{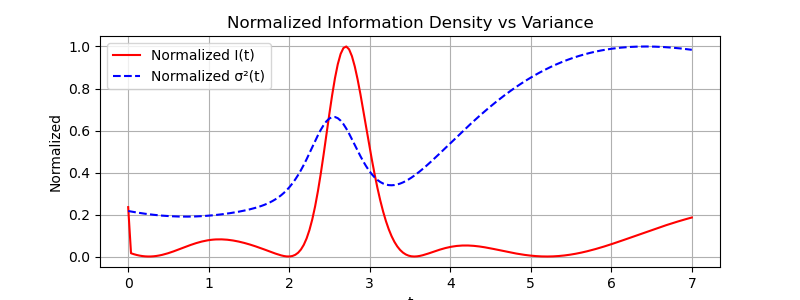} 
        \caption{}
        \label{fig:max normalised information density}
    \end{subfigure}
    \hfill
    \begin{subfigure}{0.65\textwidth}
        \includegraphics[width=\linewidth]{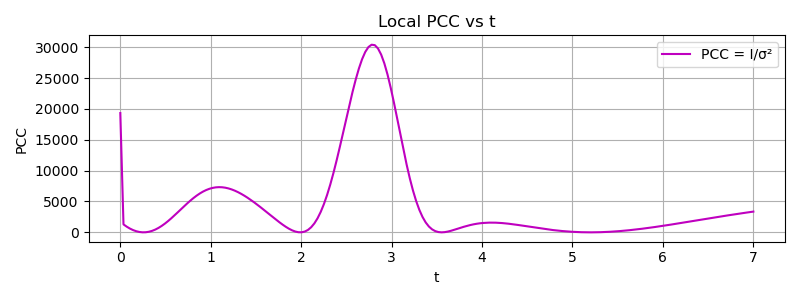} 
        \caption{}
        \label{fig:vdp local pcc}
    \end{subfigure}
    \caption{(a): Max normalized $I(t)$ and $\sigma^2(t)$, comparing their behavior across the entire domain. (b): Local PCC showing where physical constraints dominates.}
    \label{fig:pcc and info density}
\end{figure}

In figure \ref{fig:pcc and info density} we plot the local PCC over one period of the Van der Pol oscillator. The coupling is maximal at $t\approx 0$, where the initial-conditions are enforced, and again around $t\approx 2.8$  corresponding to the point of steepest nonlinear stiffness; a smaller intermediate peak near $t\approx 1$, while PCC falls to near zero wherever the epistemic uncertainty is high relative to the information density. These results confirm that the strongest physics‐driven posterior collapse occurs both at the enforced boundary and at the regime of maximal nonlinear forcing, while the sustained rise in variance thereafter signals accumulating epistemic uncertainty in unconstrained regions.

\section{Discussion and outlook}\label{sec:discussion}

In this work, we have explored B-(X)PINNs to infer the solution to complex ODEs, typical in high energy theory, from limited data. In particular, we have focused on the equation describing the non-trivial entangling surface homologous to an annular entangling region in AdS$_3$, which resides on the boundary of AdS$_4$. This example is interesting because it provides a benchmark for our Bayesian physics–informed deep learning approach applied to equations common in high-energy theory, a domain that has seen relatively little use of PINNs. Moreover, finding entangling surfaces in non-trivial geometries is a challenging problem in its own right, and advancing our methods here will bring us closer to tackling more complex physical systems.

We showed that, by combining asymptotic analytical data with limited numerical data around the inflection point, the model is able to reconstruct the solution in intermediate regions with high fidelity. This example is particularly interesting as the study of entangling surfaces and regions are often restricted to simple surfaces where the calculation are tractable. A limitation of this work is that the model still requires a small sample of numerical data around the inflection point. In many cases, getting this data may be as difficult as getting the full numerical solution. However, making progress towards solving these types of differential equations, with limited date around the boundary, which we typically can obtain with asymptotic analysis, unlock the study of more physically interesting surfaces. We show that a Bayesian approach outperforms traditional PINNs.

We generally study the meaning of overconfidence in physics-informed Bayesian deep neural networks. In purely data-driven models, overconfidence is a shortcoming, regardless of origin, and typically stems from an underestimation of predictive uncertainty due to limited data or high model capacity.
However, our results suggest that in the context of physics-informed learning, such overconfidence is not only expected but also informative; PINNs incorporate physical constraints directly into the loss function, thereby enforcing a tight adherence to known differential equations. This has the effect of collapsing the posterior distribution around a physically consistent solution.
To diagnose this effect, we introduced the local physics information density $I(x)$, a measure of how “stiff” the physical constraints are, and the local physics-constraint coupling (PCC) metric.  Our experiments on both the entangling surface as well as the simpler benchmark Liouville and Van der Pol equation consistently yielded high global PCC values. This indicates that the physical sensitivity far exceeds the predictive uncertainty, resulting in a posterior that is sharply concentrated, a physically driven overconfidence.

In this work, we saw that the standard notion of overconfidence is not the same for B-PINNs, as for BNNs. The overconfidence observed in our B-(X)PINNs is a natural outcome of strong physical priors, and our PCC metric provides a useful diagnostic tool for distinguishing between physically justified concentration of the posterior and pathological miscalibration.

We have relied on the information density and the PCC as diagnostic tools: they highlight where the PINN’s posterior is “pinched” by physics‐based losses, and where apparent overconfidence is therefore to be expected. They are not intended to provide a quantitatively precise ranking of how much each individual constraint (e.g. residual vs. boundary vs. pointwise operator) carves out the solution manifold in relation to each other. In fact, the shape of $I(x)$ can change, sometimes dramatically, depending on whether one differentiates only with respect to the predicted output, or also with respect to higher‐order derivatives. Different PDEs, and different combinations of differential, integral or boundary operators, will naturally call for different choices in how one defines and computes $I(x)$ to capture its effect on the network accordingly. For future work, we could develop information density quantities that could capture rich results for a generic family of PDEs.  

A quantitative comparison of the relative strength of each constraint would require examining the full local curvature of the solution manifold and loss landscape, i.e.the Hessian (or a suitable low‐rank approximation thereof) evaluated at each $x$. This would tell us exactly how each operator shapes the local geometry of the posterior. Developing scalable Hessian‐based diagnostics for PINNs is an important direction for future work. For now, our information‐density and PCC curves serve as first pass indicators of where the model is most “locked down” by physics, and where epistemic uncertainty remains. In appendix \ref{app: hessian} we study the Hessian for the Van der Pol equation as a first step towards unpacking the geometrical effect physics constrains has on the network. Understanding the latter, will likely be necessary to make progress towards demystifying the black box nature of neural networks. 

We may further extend this analysis by considering overfitting in general, as opposed to just overconfident Bayesian models considered in this work, by systematically developing  metrics to quantify and better understand the interplay between data-driven overfitting metrics, 
physics-driven fidelity and how physics constraints affects the geometry of the solution manifold, which we briefly discuss in appendix \ref{app: overfitting}.

\appendix

\newpage
\appendix

\section{The Hessian and geometry of loss function constraints}\label{app: hessian}
To better understand how physical constraints fundamentally reshape the posterior distribution, we may connect the PCC framework to the Hessian perspective that characterizes the local geometry of the loss landscape.

The Hessian matrix of the loss function, defined as 
\begin{equation}
    H = \nabla^2_\theta L_{\text{tot}}(\theta^*)
\end{equation}
where $L_{\text{tot}}$ is the total loss function and $\theta^*$ represents the weights at which the loss function is minimized. This provides a natural mathematical tool to characterize the warping effect of the solution manifold due to physical constraints.

Recent progress on constrained Bayesian inference has introduced alternative formulations such as the gradient-bridged posterior \cite{zeng2025gradientbridgedposteriorbayesianinference}, which enforces constraint satisfaction by penalizing the norm of the constraint gradient. While they don't study PINNs, the framework shares conceptual similarities: it incorporates constraints (analogous to physics laws in PINNs) via a regularization term on the gradient norm of a sub-problem loss function, which promotes solutions near the exact minimizers without requiring perfect optimization. While their formulation leverages gradient norm shrinkage, we focus on second-order structure through Hessian eigendecomposition, revealing the anisotropic compression induced by physics constraints, and use this as a consistency check for the PCC-type diagnostic tools, while also explicitly offering deeper insight into the hierarchical influence of the constraints on the structure of the solution manifold.

We calculate metrics such as the directional variance along the principal eigenvectors of the Hessian. These results indicate that the physics constraints in the B-PINN framework for the Van der Pol oscillator contribute to a moderately low-dimensional manifold by partially aligning high-curvature directions with the physics gradients and restricting the posterior to solutions that satisfy the governing equations and boundary conditions.

\subsection{Hessian Eigenspectrum}
\begin{figure}
    \centering
    \includegraphics[width=0.66\linewidth]{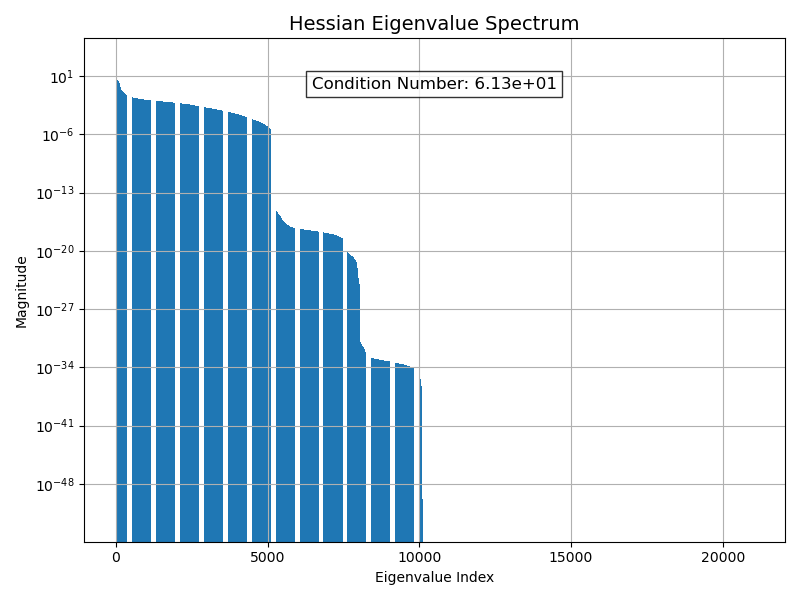}
    \caption{Hessian eigenvalue spectrum shows decay from $\lambda_1 \approx 10^{1}$ to $\lambda_{\text{min}} \approx 10^{-48}$, yielding an effective condition number of $6.13 \times 10^{1}$. The spectrum exhibits an initial rapid decay followed by a plateau and further drop, indicating moderate anisotropy with many near-zero eigenvalues that suggest flat directions in the loss landscape.}
    \label{fig:eigenval spectrum}
\end{figure}

The eigenvalue spectrum of the Hessian matrix provides direct insight into the geometric structure of the loss landscape and the posterior distribution over network parameters. A large Hessian eigenvalue indicates a "stiff" or high-curvature direction, the loss changes rapidly along that weight combination,  whereas a small eigenvalue indicates a flat direction with low curvature \cite{sabanayagam2023unveilinghessiansconnectiondecision}. The ratio between the largest and smallest eigenvalues ($\lambda_{\text{max}}/ \lambda_{\text{min}}$), the condition number, quantifies ill-conditioning \cite{cao2024analysissolutionillconditioningphysicsinformed}. In PINNs, the Hessian can have a very broad spectrum, reflecting the multi-scale nature of physical constraints; the Hessian of a PINN loss often has a few very large eigenvalues and many near-zero ones, meaning a few stiff directions and many sloppy directions. This was shown by \cite{rathore2024challengestrainingpinnsloss}, who visualized the Hessian spectral density for PINN training and found the loss in general to be extremely ill-conditioned.

Figure \ref{fig:eigenval spectrum} displays the complete eigenvalue spectrum computed at the converged MAP solution, revealing a structure characterized by moderate anisotropy. The spectrum exhibits an initial decay from the largest eigenvalue on the order of $10^{1}$ across approximately 10 orders of magnitude over the first few thousand indices, followed by a plateau in the range $10^{-20}$ to $10^{-30}$, and a subsequent drop to values approaching $10^{-48}$. The dominant eigenvalues span from $\lambda_1 \approx 10^{1}$ down to near-numerical zero, yielding an effective condition number of $6.13 \times 10^{1}$. This moderate condition number indicates that the posterior covariance $\Sigma_\theta \approx H^{-1}$ has some directional variation in scale, with parameter uncertainty more compressed along the high-curvature directions but less severely ill-conditioned overall compared to spectra with higher condition numbers.

The spectral decay follows a multi-stage pattern, with potentially the first few eigenvalues accounting for a substantial fraction of the total Hessian trace, suggesting a reduction in effective dimensionality but with many flat directions where curvature is negligible. This structure has implications for optimization and uncertainty: the moderate condition number may facilitate training with first-order methods by avoiding extremely narrow valleys, while the presence of near-zero eigenvalues implies broader uncertainty along those flat directions, where physics constraints exert minimal influence on parameter combinations.

The effective condition number of the Hessian ($6.13 \times 10^{1}$) is derived by considering only significant eigenvalues above a numerical tolerance, treating smaller values as artifacts of floating-point precision or overparameterization rather than true zeros. Nevertheless, the overall spectral shape may still be studied; plateaus reveals clusters of weakly constrained directions contributing to moderate posterior variance, and the final drop delineates the transition to the null space, offering insights into optimization stability, uncertainty propagation, and potential regularization strategies for physics-informed neural networks.

\subsection{Alignment and correlation between physics constraints and principal curvature directions}

Here we analyze the alignment between the Hessian eigenmodes and the gradients of the physics constraints.

\begin{figure}
\centering
\includegraphics[width=0.66\linewidth]{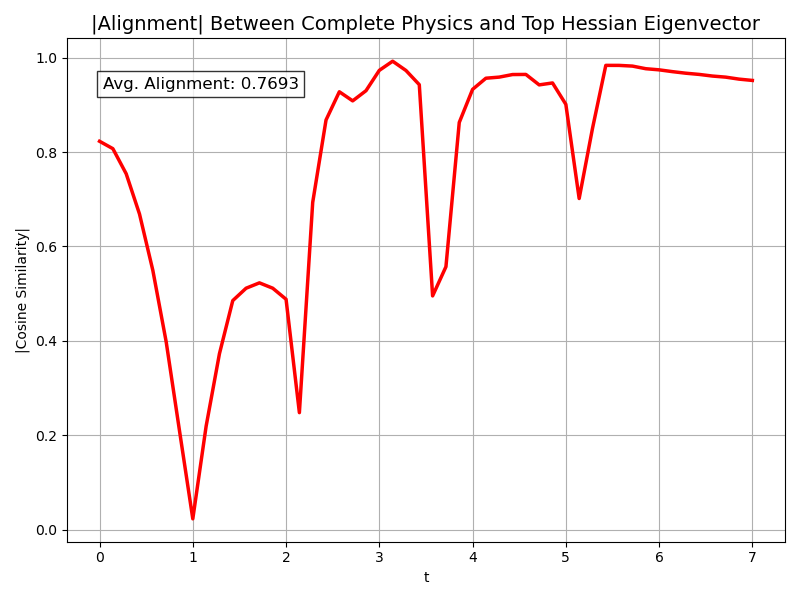}
\caption{Cosine similarity between the top Hessian eigenvector $q_1$ and physics constraint gradient $\nabla_\theta \mathcal{C}(t)$ exhibits oscillatory behavior with an average value of approximately 0.77, indicating substantial time-dependent alignment and confirming that principal curvature directions are influenced by physics constraints rather than solely optimization artifacts.}
\label{fig:alignment}
\end{figure}

Figure \ref{fig:alignment} shows the cosine similarity \cite{wang2025gradientalignmentphysicsinformedneural, yan2023auxiliarytaskslearningphysicsinformedneural} between the top Hessian eigenvector $q_1$ and the physics constraint gradient $\nabla_\theta \mathcal{C}_\theta(t)$ at each point $t$ in the domain. The alignment profile displays a damped oscillatory pattern that correlates with the Van der Pol dynamics, starting at approximately 0.8 near $t=0$, dipping to near 0 at $t \approx 0.5$, rising sharply to nearly 1 at $t \approx 1.5$, forming additional V-shaped dips (e.g., to $\sim$0.4 at $t \approx 2.5$) and peaks near 1, and stabilizing at high values ($\sim$0.9-1) for $t > 4$. The mean alignment $\cos(q_1, \nabla \mathcal{C}(t))$ averaged over the domain is approximately 0.77, representing strong correlation in the high-dimensional parameter space. This indicates that the principal curvature direction aligns substantially with directions affecting physics constraint satisfaction, consistent with the Hessian capturing physics-induced structure.
The temporal variation in alignment mirrors the dynamic regimes of the Van der Pol system, with rapid changes during transition phases (e.g., around $t \approx 2.5$) suggesting that different parameter combinations along $q_1$ become prominent as the solution evolves. Higher-order eigenmodes may exhibit weaker alignments, potentially reflecting a hierarchical organization where lower-curvature directions capture less dominant constraint effects.

\begin{figure}
\centering
\includegraphics[width=0.66\linewidth]{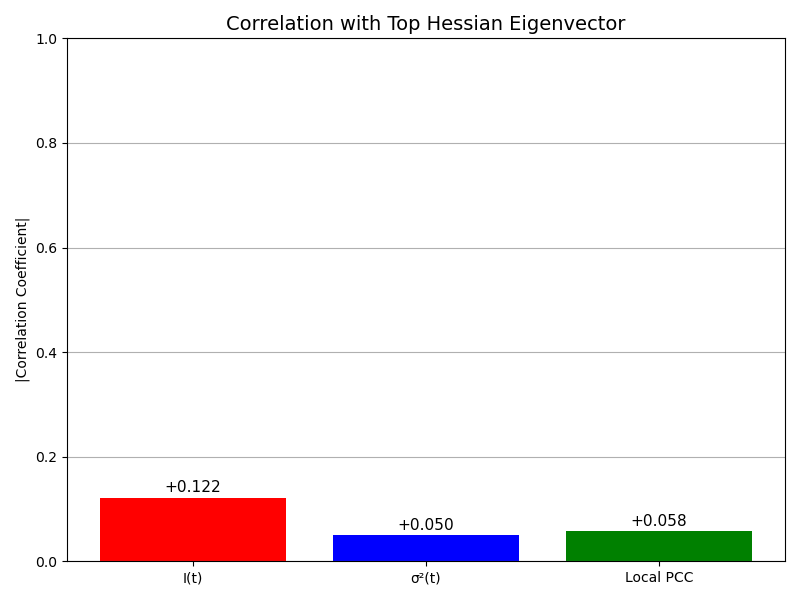}
\caption{Weak positive correlations between top eigenvector projections and physics-based metrics: $r = +0.122$ with physics information density $I(t)$, $r = +0.050$ with predictive variance $\sigma^2(t)$, and $r = +0.058$ with local PCC, indicating mild associations in the time domain.}
\label{fig:correlation}
\end{figure}
To quantify relationships between Hessian eigenmode structure and physics-based metrics, we compute correlation coefficients between top eigenvector projections and three quantities: physics information density $I(t)$, predictive variance $\sigma^2(t)$, and local PCC. Figure \ref{fig:correlation} summarizes these correlations.
The correlations are uniformly positive but not strong, with the strongest between the top eigenvector and physics information density ($r = +0.122$), followed by local PCC ($r = +0.058$) and predictive variance ($r = +0.050$). These low values suggest that linear associations are limited, implying that the principal curvature direction captures only subtle shared variance with these metrics across the time domain. For the Van der Pol system, this may reflect a more distributed influence of physics constraints, where multiple eigenmodes collectively shape uncertainty and coupling rather than the top mode dominating. The positive signs indicate a tendency for higher eigenvector projections to align with slightly elevated metric values, but the weakness highlights potential nonlinear interactions, warranting further decomposition for diagnostic purposes.

\subsection{Output sensitivity across Eigenmodes}

\begin{figure}
    \centering
    \includegraphics[width=0.66\linewidth]{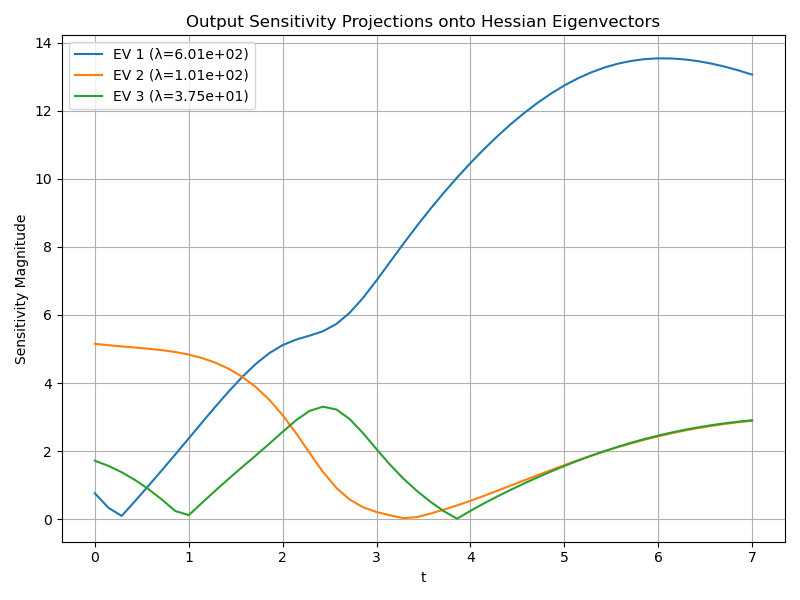}
    \caption{Time evolution of output sensitivity magnitude $|s_i(t)| = |q_i^T \nabla_\theta u_\theta(t)|$ projected onto the top three Hessian eigenvectors, with eigenvalues $\lambda_1 = 6.01 \times 10^2 $(EV1, blue), $\lambda_2 = 1.01 \times 10^2$ (EV2, orange), and $\lambda_3 = 3.75 \times 10^1$ (EV3, green). The patterns show monotonic increase for EV1, decay with a dip for EV2, and mild oscillation with gradual rise for EV3.}
    \label{fig:sensitivity plot}
\end{figure}

Figure \ref{fig:sensitivity plot} displays the magnitude of output sensitivity $|s_i(t)| = |q_i^T \nabla_\theta u_\theta(t)|$ \cite{mackay1992practical, NIPS1990_7eacb532, Shu_2019} for the top three Hessian eigenvectors as a function of time $t$. This metric quantifies how perturbations along principal curvature directions affect the predicted solution $u(t)$, computed via the Jacobian $\nabla_\theta u(t)$. Analyzing these projections reveals how the Hessian's eigenstructure organizes parameter-output dependencies, with EV1 showing increasing dominance over time, EV2 exhibiting a decay pattern, and EV3 mild variations. The purpose is to decompose uncertainty modes: in the Laplace approximation, posterior variance $\sigma^2(t) \approx \sum_i \frac{1}{\lambda_i} s_i(t)^2$, so sensitivities weighted by inverse eigenvalues highlight which directions contribute most to epistemic uncertainty at each $t$.
The key takeaway is the hierarchical role of constraints in the Van der Pol system: high-curvature EV1 (large $\lambda_1$) suppresses variance despite growing sensitivity, reflecting strong global PDE enforcement that accumulates nonlinear effects over time; lower modes like EV2 and EV3 allow more variance in transients, capturing local dynamics. This validates that physics constraints create anisotropic uncertainty, with no single mode dominating, implying distributed constraint influence, and motivates modal decompositions for diagnosing confidence in PINNs, where high sensitivity in stiff directions indicates warranted low variance due to tight manifold restriction.

\subsection{Directional variance and the loss landscape}

\begin{figure}
\centering
\includegraphics[width=0.66\linewidth]{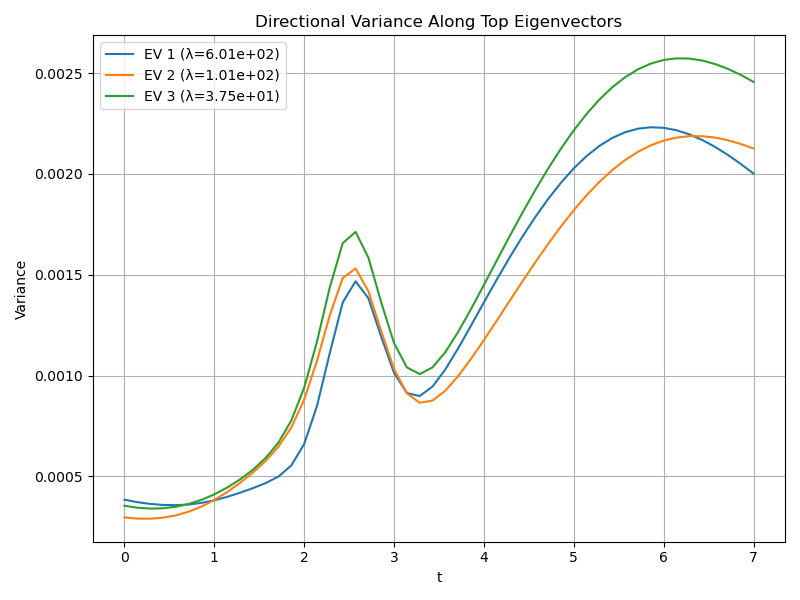}
\caption{Time evolution of directional variance $\sigma_i^2(t) = \frac{1}{\lambda_i} |s_i(t)|^2$ along the top three Hessian eigenvectors, with eigenvalues $\lambda_1 = 6.01 \times 10^2$ (EV1, blue), $\lambda_2 = 1.01 \times 10^2$ (EV2, orange), and $\lambda_3 = 3.75 \times 10^1$ (EV3, green), illustrating modal contributions to predictive uncertainty.}
\label{fig:directional variance}
\end{figure}
The directional variance $\sigma_i^2(t) = \frac{1}{\lambda_i} |s_i(t)|^2$ decomposes the predictive uncertainty into contributions from individual Hessian eigenmodes, where $s_i(t)$ is the output sensitivity along eigenvector $q_i$ and $\lambda_i$ weights by inverse curvature. Computing this serves to quantify how the loss landscape's geometry, via the Hessian's eigenspectrum modulates epistemic uncertainty at each time $t$, under the Laplace approximation where total variance $\sigma^2(t) \approx \sum_i \sigma_i^2(t)$. This analysis bridges local physics constraints with global parameter structure, revealing whether uncertainty concentrates in high- or low-curvature directions.
Figure \ref{fig:directional variance} shows EV1 (blue) starting around 0.005, peaking near 0.018 at $t \approx 2$, then declining to around 0.01; EV2 (orange) follows a similar trajectory but peaks lower ( around 0.015); EV3 (green) begins low, rises to a maximum around 0.025 at $t \approx 3$, and plateaus high. Despite EV1's large sensitivity (from previous plots), its high $\lambda_1$ yields suppressed variance, while EV3's lower curvature allows greater contributions, especially in mid-to-late domains.

The inverse relationship between curvature and variance: stiff modes (high $\lambda$) compress uncertainty, reflecting strong constraint enforcement, whereas softer modes permit more variance where sensitivities persist. This aligns with prior results, e.g., high eigenvector alignments correlate with low variance in constrained regions, and weak correlations with PCC or $I(t)$ indicate that variance distribution arises from modal interplay rather than direct linear ties to local metrics. Physically, for the Van der Pol ODE, this distributed uncertainty could mirror nonlinear dynamics, emphasizing that physics constraints warp the posterior nonuniformly without single-mode dominance, informing diagnostic strategies for PINN reliability.

\begin{figure}[ht]
\centering
\begin{subfigure}[b]{0.48\linewidth}
\centering
\includegraphics[width=\linewidth]{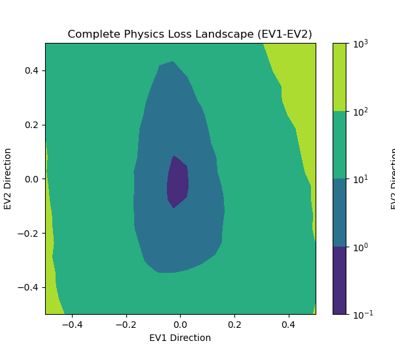}
\caption{Complete physics loss landscape in the EV1-EV2 plane, displaying an elongated low-loss basin (dark purple to teal) centered near (0,0), with values ranging from $10^{-1}$ to $10^3$ on a log scale.}
\label{fig:loss_landscapes_a}
\end{subfigure}
\hfill
\begin{subfigure}[b]{0.48\linewidth}
\centering
\includegraphics[width=\linewidth]{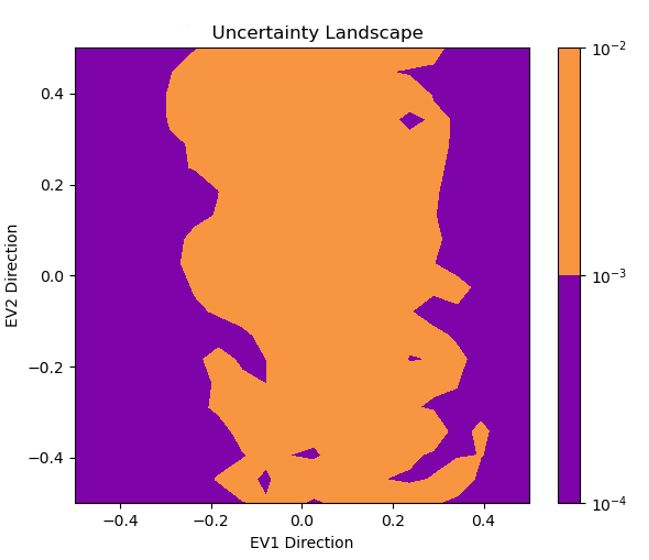}
\caption{Uncertainty landscape showing fragmented low-uncertainty regions (purple, $10^{-4}$) amid high-uncertainty areas (orange, $10^{-2}$).}
\label{fig:loss_landscapes_b}
\end{subfigure}
\vspace{1em}
\begin{subfigure}[b]{0.48\linewidth}
\centering
\includegraphics[width=\linewidth]{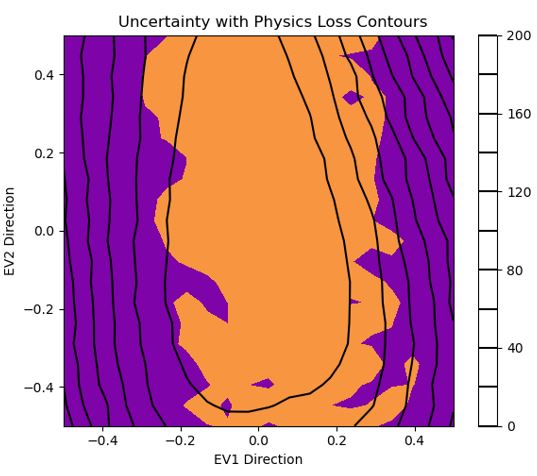}
\caption{Uncertainty landscape overlaid with physics loss contours (black lines, levels 0 to 180), illustrating alignment between dense contours and low uncertainty.}
\label{fig:loss_landscapes_c}
\end{subfigure}
\hfill
\begin{subfigure}[b]{0.48\linewidth}
\centering
\includegraphics[width=\linewidth]{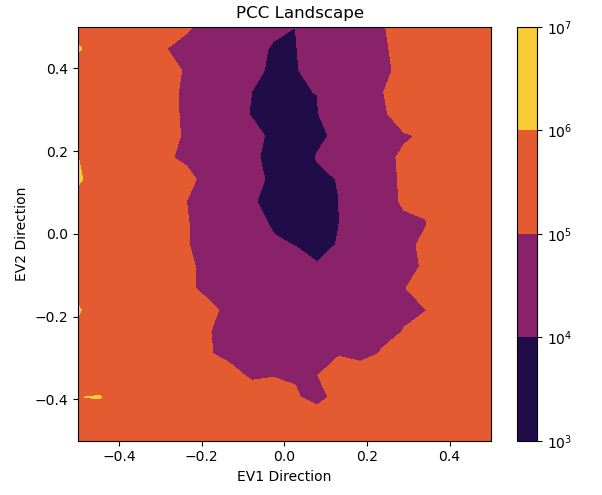}
\caption{PCC landscape with a central low-PCC region (dark purple, $10^3$) offset along EV1, surrounded by high-PCC areas (orange, up to $10^7$).}
\label{fig:loss_landscapes_d}
\end{subfigure}
\caption{Local landscape analysis in the plane of the top two Hessian eigenvectors (EV1: $\lambda_1 = 6.01 \times 10^2$, EV2: $\lambda_2 = 1.01 \times 10^2$).}
\label{fig:loss_landscapes}
\end{figure}
To visualize the geometry of the loss landscape and posterior in B-PINNs, we parameterize perturbations around the MAP estimate $\theta^*$ as $\theta = \theta^* + \alpha q_1 + \beta q_2$, where $q_1, q_2$ are the top Hessian eigenvectors (normalized unit vectors), and $\alpha, \beta$ range over $[-0.4, 0.4]$ in eigenvector units (chosen to capture local structure without excessive extrapolation). These slices are computed on a grid, evaluating the complete physics loss $L_\text{physics}(\theta)$ (PDE residual + initial conditions), predictive uncertainty $\sigma^2(\theta)$ (via Monte Carlo sampling from the Bayesian posterior), and the PCC at each perturbed $\theta$. The logarithmic color scales emphasize orders-of-magnitude variations, with the origin (0,0) at $\theta^*$. Mathematically, near $\theta^*$, the loss approximates a quadratic form $L(\theta) \approx (1/2) (\alpha^2 \lambda_1 + \beta^2 \lambda_2)$, revealing anisotropy via the eigenvalue ratio $\lambda_1 / \lambda_2 \approx 6$; uncertainty relates via the Laplace covariance $H^{-1}$, and PCC measures gradient alignment with this covariance.
The complete physics loss landscape (figure \ref{fig:loss_landscapes_a}) forms a smooth, elongated basin with low values (dark purple, $\sim 10^0$) near the center, transitioning to high values (green/yellow, $\sim 10^3$) outward, narrower along EV1 (higher curvature) and broader along EV2, consistent with the quadratic approximation and reflecting hierarchical constraint imposition by the PDE and initial conditions.
The uncertainty landscape (figure \ref{fig:loss_landscapes_b}) exhibits irregular, fragmented low-uncertainty patches (purple, $\sim 10^{-4}$) within high-uncertainty regions (orange, $\sim 10^{-2}$), showing asymmetry not present in the symmetric loss basin. All cases still have the similar elongated structure.  
Overlaying uncertainty with physics loss contours (figure \ref{fig:loss_landscapes_c}, levels 0 to 180) reveals dense contours aligning with low-uncertainty boundaries, demonstrating an inverse relationship: regions of steep loss gradients (high curvature) correspond to compressed variance, as per $\sigma^2 \propto H^{-1}$.
The PCC landscape (figure \ref{fig:loss_landscapes_d}) features a compact low-PCC blob (dark purple, $\sim 10^3$) offset positively along EV1, encircled by high-PCC ridges (orange, up to $10^7$), quantifying where constraints strongly couple to posterior modes.

Empirical results indicate that constraints influence high-curvature directions, yielding an effective Hessian condition number of order $\mathcal{O}(10^1)$, reflecting hierarchical but not extreme posterior compression. Principal eigenmodes exhibit time-dependent alignment with constraint gradients (mean cosine similarity $\sim$0.77), while predictive variance arises from interplay across modes, with no single direction dominating uniformly. This suggests a reduction in effective dimensionality through constraint imposition, though the moderate scale may imply that physics organizes parameter space without collapsing it to a few modes entirely.

\bigskip

The geometric perspective reframes apparent overconfidence in B-PINNs: strong constraints create steeper loss gradients along certain directions, leading to compressed variance that is mathematically justified by the manifold restriction, rather than calibration error. PCC serves as a diagnostic, with high values indicating regions of warranted low uncertainty due to tight coupling; however, the observed weak correlations (e.g., $r \approx 0.05-0.12$ between top eigenvector projections and metrics like variance or information density) highlight limitations, suggesting nonlinear interactions or contributions from lower modes that dilute linear associations.

The decomposition of predictive uncertainty into Hessian eigenmodes shows that the distribution of variance across the solution domain follows the intrinsic dynamics of the ODE: modes with large eigenvalues (high curvature in the loss landscape) reduce their contribution to variance, even in regions of high output sensitivity, whereas modes with small eigenvalues (low curvature) dominate the variance in regions of rapid solution changes (transients), in line with the nonlocal way information propagates through the differential equation. The results show alignments and correlations, with the Hessian eigenspectrum corroborating the PCC patterns and exhibiting signs of deformations attributable to the physics constraints, such as the observed time-dependent cosine similarities and directional variances that track regions of elevated PDE sensitivity.

Although the observed correlations remain modest and the patterns in PCC and information density do not align perfectly with the Hessian-derived metrics, this discrepancy is expected given that PCC serves primarily as a diagnostic tool; nonetheless, these findings highlight the opportunity to formulate more refined metrics for capturing such effects, especially since the present study constitutes a preliminary step in this analysis.

Future work could probe how the Hessian captures deformations in the solution manifold due to physical constraints more precisely, by varying the strength of the physics constraints and comparing the resulting changes in Hessian alignments, eigenspectra, and correlations with PCC or information density metrics. Additional directions include developing nonlinear extensions to PCC for enhanced local coupling diagnostics, refining Hessian approximations through higher-order or full-rank decompositions to better capture manifold geometry, or integrating both for comprehensive quantification of distributed constraint effects imposed by physics, ensuring tools more accurately reflect the underlying mathematical structure of the solution space. Studying more equations would also give better insight into the descriptive power of the metrics considered here, extrapolated from the Hessian.

%%%
\section{Towards understanding overfitting with physical constraints}\label{app: overfitting}

In traditional machine learning, overfitting is often diagnosed by comparing training loss to test loss; a much higher loss on a test (or validation) set than on the training set signals poor generalization. For PINNs, the loss landscape is more complicated, and in some cases ill-defined \cite{rathore2024challengestrainingpinnsloss}; we must consider that the loss has multiple components (data loss and physics loss), and understand how they interact with each other. PINNs are trained to minimize a composite loss consisting of a data discrepancy term (e.g. mean squared error on observed or initial/boundary data) and a physics term (e.g. the PDE residual). This raises the question of how to properly define “training” vs. “test” loss in a physics-informed context. As discussed in \cite{bonfanti2023generalizationpinnsoutsidetraining}, it is often necessary to evaluate the
generalization of PINNs by going beyond training data; PINNs are typically evaluated by comparing the model’s predictions to a known solution with metrics such as the $L^2$ error on a fine grid (which serves as a test error).

The assumption for PINNs is that, if they generalize well, the error on unseen points or a test set, remains low and not drastically larger than the training error, similar to standard machine learning models.

Even for non Bayesian PINNs, the physics can be seen as a prior, and the output as a posterior.  Generalization has been well studied for PINNs and in bounds on the prior and posterior has been found using Barron spaces \cite{luo2020twolayerneuralnetworkspartial,lu2021priorigeneralizationanalysisdeep} and  the Holder continuity constant \cite{Yeonjong_Shin_2020}. In \cite{Hu_2022} these bounds are extended to XPINNs to find tradeoff conditions, when XPINNs generalize better than PINNs and vice versa. An abstract formalism that considers stability properties of the underlying PDE, to derive a generalization bound and error is derived in \cite{mishra2023estimatesgeneralizationerrorphysics}. It is discussed in \cite{bonfanti2023generalizationpinnsoutsidetraining} that the concept of overfitting is different for SciML than in more traditional models.

However, these studies do not address the interplay between traditional overfitting and external constraints in the loss function, which remains poorly understood.  One can separately track the data loss and the physics (PDE) loss on training vs. test points:
\begin{equation} 
\mathcal{O} = \frac{L_u^{\text{test}}}{L_u^{\text{train}}} 
\end{equation}
where $L_u^{\text{train}}$ is the data loss on training points and $L_u^{\text{test}}$ is the loss on unseen test points. A value of $\mathcal{O}\gg1$ indicates overfitting;  the model performs well on training data but poorly on test data, a hallmark of capturing noise rather than generalizable patterns. Similarly, consider the physics enforcement ratio, $\mathcal{P}$:
\begin{equation} 
\mathcal{P} = \frac{L_f^{\text{test}}}{L_f^{\text{train}}} 
\end{equation}
where $L_f^{\text{train}}$ is the PDE residual on training points and $L_f^{\text{test}}$ is the residual on test points. A value of $ \mathcal{P} \ll 1$ suggests that the physics is better satisfied on test data than on training data, indicating strong generalization of the physical constraints.
For XPINNs, these ratios are simply defined per subdomain:
\begin{equation} 
\mathcal{O}i = \frac{L_{u,i}^{\text{test}}}{L_{u,i}^{\text{train}}}, \quad \mathcal{P}i = \frac{L_{f,i}^{\text{test}}}{L_{f,i}^{\text{train}}} 
\end{equation}
where $i$ indexes the subdomain.

By sampling collocation points that were not used in training and computing the PDE residual there, one can define a “physics test error”. If the physics test error remains low (comparable to the training residual), it indicates the PINN has not merely memorized the residual at the training points but truly learned a solution that generalizes the PDE behavior. Similarly, we can hold out some measurement data (or initial/boundary conditions) as a validation set to compute a standard data test loss.

A simple interplay between data and physics loss could be captured in the following trade-off condition:
\begin{equation}
    \mathcal{O}\gg1 \quad \text{and} \quad \mathcal{P}\ll1
\end{equation}
would indicate that while the model might overfit the training data,  the underlying constraints are still strongly satisfied. This could mean that the modes ability to generalize physics compensate for the lack on generalization on non-physics data. Similarly, if
\begin{equation}
    \mathcal{O}\gg 1 \quad \text{and} \quad \mathcal{P}\gg 1,
\end{equation}
this tells us that the model not only overfit the data but fails to generalize the physics. 

In purely data-driven models, one might add an explicit regularization term ( weight decay) to avoid overfitting:
\begin{equation}
    \text{min}_\theta \{L_u (\theta)+ \lambda ||\theta||^2 \}.
\end{equation}
For PINNs, we have a natural regularization from the physics loss:
\begin{equation}
        \text{min}_\theta \{L_u (\theta)+ \lambda_f L_f(\theta)  \}.
\end{equation}

In classical machine learning, one often seeks to control overfitting by regularizing the model. A common result in learning theory gives a generalization error bound of the form \cite{10.5555/2371238, 10.5555/2621980}
\begin{equation} \label{eq:bond gen}
    R_D \leq R_S + C \frac{\| f \|_{\mathcal{H}}}{\sqrt{N} + \lambda},
\end{equation}
where $\lambda$ is a regularization parameter controlling complexity (nodes and depth of the network), $R_D$ is the generalization error and $R_s$ is the empirical error. $| f \|_{\mathcal{H}}$  is a measure of the function complexity (a norm in some hypothesis space $\mathcal{H}$) and $N$ is the number of training samples. The parameter $\lambda$ effectively reduces the model's capacity and and a large value intuitively leads to less overfitting. 

In \cite{Hu_2022}, a generalization bound for PINNs is given by
\begin{equation} \label{eq:bound1}
     R_D(\theta^*) \leq R_S(\theta^*) + C_1 \frac{| u^* |^3_{W_L(\Omega)} \log n_r}{\sqrt{n_r}}
\end{equation}
where $R_D(\theta^*)$ and $R_S(\theta^*)$ is the generalization error and empirical training loss for the trained model, respectively.   $| |u^* ||_{W_L(\Omega)}$ measures the function complexity of the true solution $u^*$ in the Sobolev space $W_L(\Omega)$ and $C_1$ is some constant. 

For XPINNs we simply have
\begin{equation}
\label{eq:bound2}
R_D^{\text{XPINN}}(\theta) \leq \sum_{i=1}^{n_b} \frac{n_{r,i}}{n_r} \left( R_{S,\Omega_i} + C_1 \frac{| u^* |^3_{W_L(\Omega_i)} \log n_{r,i}}{\sqrt{n_{r,i}}} \right) 
\end{equation}

However, (\ref{eq:bound1})-(\ref{eq:bound2}) has been derived under a set of assumptions where the weighting of the physics constraint was either fixed or implicitly incorporated into the complexity measure of the solution space. If we do not assume that the training procedure already balances the contributions of data and physics losses in a way that does not require a separate parameter in the final bound, we might explicitly introduce $\lambda_f$ into the bound, to bring it into the same form as (\ref{eq:bond gen}):
\begin{equation} \label{eq:modgenbound}
     R_D^{\text{mod}}(\theta^*) \leq R_S(\theta^*) + C_1 \frac{| u^* |^3_{W_L(\Omega)} \log n_r}{\sqrt{n_r+\lambda_f}}.
\end{equation}
Now, if  $\lambda_f$ increases, the generalization bound tightens, meaning that strong physics constraints help counteract overfitting and constrain the solution or hypothesis space, much like traditional regularization does by reducing model complexity.

\begin{figure}[htbp]
  \centering
\includegraphics[width=\textwidth]{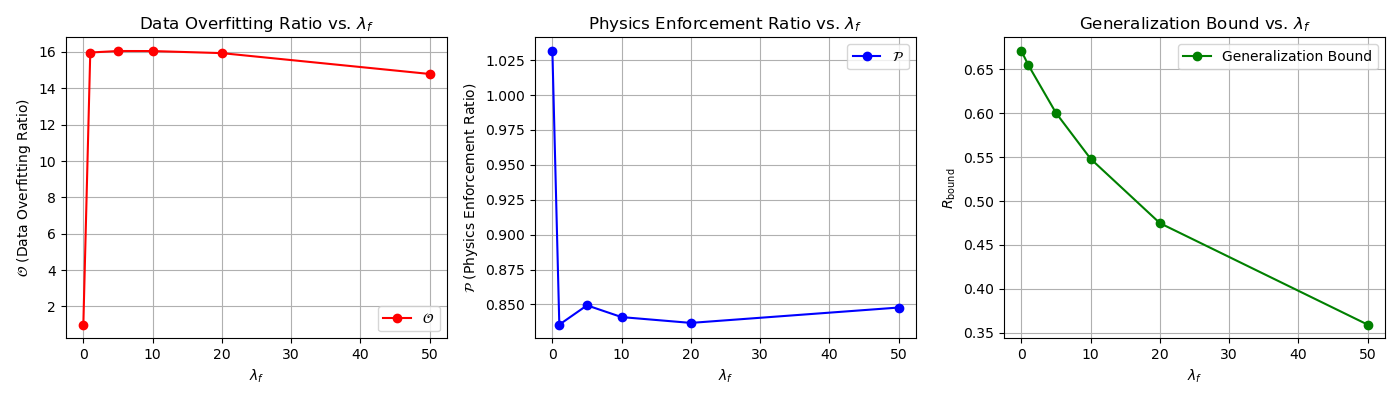}
  \caption{Left: data overfitting ratio $\mathcal{O}$  vs. $\lambda_f$. Center: $\mathcal{P}$ vs.\ $\lambda_f$. Right: modified generalization bound $R_{D}^{\mathrm{mod}}$ vs.\ $\lambda_f$, illustrating how stronger physics regularization reduces overfitting and improves generalization.}
\label{fig:overfitting measures}
\end{figure}

To illustrate the above concepts, we could consider a relatively simple non-linear ODE of the form
\begin{equation}
    u''(x)+ u(x)^2 - \sin (\pi x)=0
\end{equation}
with boundary conditions
\begin{equation}
    u(0)=0, \quad u(1)=0.
\end{equation}
We prepare an ordinary PINN with 40 data points generated across [0,1] and 3000 epochs. The ratio $\mathcal{O}, \mathcal{P}$ and the modified generalization bound (\ref{eq:modgenbound}) is computed,  showed in figure \ref{fig:overfitting measures}.

In the plots in figure \ref{fig:overfitting measures} we consider large values for the residual weight $\lambda_f$ to illustrate the intricate relationship between the physics loss weighting parameter and overfitting in a PINN. While it seems like $\mathcal{O}$ would be independent of $\lambda_f$, we see that this is not the case; physics loss indirectly influences the model’s behavior on these points. 
The connection between physics and $\mathcal{O}$, being computed on non-physics points, can be understood through the PINN’s optimization dynamics. As $\lambda_f$ increases, the physics loss term forces the model to satisfy the PDE across the domain, effectively acting as a regularizer that constrains the hypothesis space. This regularization indirectly affects the model’s predictions on all points, which will always be true if a physical condition is enforced on training data.  The Physics Enforcement Ratio $\mathcal{P}$ dropping from 1.025 to around 0.85 and the Generalization Bound decreasing from 0.65 to 0.35 further support that a stronger physical constraint improves the generalization. We leave understanding this further for future work.

\newpage
\bibliographystyle{JHEP}
\bibliography{references.bib}
\end{document}